# Online Scaling of NFV Service Chains across Geo-distributed Datacenters

Yongzheng Jia, Chuan Wu, Zongpeng Li, Franck Le, and Alex Liu

*Abstract*—Network Function Virtualization (NFV) is an emerging paradigm that turns hardware-dependent implementation of network functions (i.e., middleboxes) into software modules running on virtualized platforms, for significant cost reduction and ease of management. Such virtual network functions (VNFs) commonly constitute service chains, to provide network services that traffic flows need to go through. Efficient deployment of VNFs for network service provisioning is key to realize the NFV goals. Existing efforts on VNF placement mostly deal with offline or one-time placement, ignoring the fundamental, dynamic deployment and scaling need of VNFs to handle practical time-varying traffic volumes. This work investigates dynamic placement of VNF service chains across geo-distributed datacenters to serve flows between dispersed source and destination pairs, for operational cost minimization of the service chain provider over the entire system span. An efficient online algorithm is proposed, which consists of two main components: (1) A regularization-based approach from online learning literature to convert the offline optimal deployment problem into a sequence of one-shot regularized problems, each to be efficiently solved in one time slot; (2) An online dependent rounding scheme to derive feasible integer solutions from the optimal fractional solutions of the one-shot problems, and to guarantee a good competitive ratio of the online algorithm over the entire time span. We verify our online algorithm with solid theoretical analysis and trace-driven simulations under realistic settings.

## I. INTRODUCTION

Traditionally, network functions such as firewalls, proxies, network address translators (NATs) and intrusion detection systems (IDSs) are implemented by dedicated hardware middleboxes, which are costly and difficult to scale. The emerging paradigm of network function virtualization (NFV) aims to revolutionize network function provisioning by running the respective software in standard virtualized platforms, *e.g.*, virtual machines (VMs) on industry-standard servers, in order to achieve fundamental flexibility in deployment and management, as well as significant cost reduction [1].

To realize these goals, a fundamental challenge is to strategically deploy the virtualized network functions (VNFs), in terms of the number of instances for each VNF and the deployment locations, and to dynamically scale the deployment, by adding/removing instances of VNFs with increase/decrease of the demand. In particular, the network functions are typically connected to compose service chains [2] that provide different network services [3][4]. Along a service chain, network traffic flows are required to go through multiple stages of network function processing in a particular order. A service chain can be placed within one datacenter (*e.g.*, the service chain "Firewall→IDS→Proxy" providing a company's access service is typically deployed in an on-premise datacenter), or distributed across multiple datacenters, (*e.g.*, for the virtualization of WAN optimizers, IP Multimedia Subsystem (IMS), mobile core networks [1]). For the example of WAN optimization, deduplication or compression functions are deployed close to the sender of a flow and traffic shaping can happen anywhere along the route from the sender to the receiver. For virtualization of a control plane service chain in an IMS [5], namely "P-CSCF→S-CSCF", instances of P-CSCF, first contact of a user for call registration, should be distributed close to geo-dispersed callers, while S-CSCF for session control can be located more in the middle between a caller and a callee.

This study focuses on geo-distributed deployment of service chains and tackles the challenges of dynamic deployment and scaling of VNF instances in the chains. We take the perspective of a network service provider, who rents VMs from cloud datacenters owned by the respective datacenter or cloud operators, deploys VNFs on the VMs and assembles service chains for the usage of its flows. For example, the network service provider can provide service chains for WAN optimization consisting of different WAN optimizers, or can be an IMS provider who seeks to virtualize its control plane and data plane service chains (an example data plane service chain in an IMS is "Firewall→IDS→Transcoder").

We target cost minimization for the network service provider by optimally deploying geo-distributed service chains. The intriguing challenges are as follows. *First*, the deployment of VMs to run different network functions is not only relevant to VM rental cost, but also decides the bandwidth (cost) needed in-between datacenters to accommodate traffic flows passing through geo-dispersed network functions, which further decides quality of the network service, *e.g.*, end-to-end delay experienced by the flows. *Second*, when multiple service chains are in place, they may well share one or multiple common network functions, and may hence exploit the same collection of VNF instances deployed; this further demands deliberate computation of the geographical deployment of the shared instances, in order to balance the end-to-end

Yongzheng Jia and Chuan Wu are with The University of Hong Kong, Hong Kong (e-mail: jiayz13@mails.tsinghua.edu.cn, cwu@cs.hku.hk).
Zongpeng Li is with University of Calgary, Canada (e-mail: zongpeng@ucalgary.ca).
Franck Le is with IBM T.J. Watson Research Center, U.S.A. (e-mail: fle@us.ibm.com).
Alex Liu is with Michigan State University, U.S.A. (e-mail: alexliu@cse.msu.edu).

2performance experienced by different flows between various sources and destinations. *Last* but most importantly, we seek to make efficient online deployment and scaling decisions on the go with the variation of flows, while guaranteeing good performance over the long run of the system.

Existing efforts on VNF placement mostly deal with the offline or one-time placement, ignoring the fundamental, dynamic deployment and scaling demand of VNF service chains to handle practical time-varying traffic flows (see Sec. II for detailed discussions). In contrast, this work investigates dynamic placement of VNF service chains in geo-distributed cloud datacenters to serve dynamically-generated flows between various source/destination pairs across the globe, for service cost and delay minimization. We show that even in the offline setting, the problem we consider renders an NP-hard combinatorial nature, leading to significant difficulty in efficient online algorithm design. Looking deep into the structure of the problem, we design an efficient online algorithm based on the state-of-the-art online learning techniques and a well-designed dependent rounding scheme. Our detailed contributions are summarized as follows.

*First*, we formulate a practical online cost minimization problem enabling dynamic deployment and removal of VNF instances in different datacenters, as well as dynamic traffic flow routing among VNF instances in the respective service chains. Various deployment and running costs of VNF instances in different datacenters are considered, in addition to time-varying bandwidth costs to transmit flows into and out of a datacenter. As a key QoS performance indicator for network service provisioning, the average end-to-end delays of the flows are also formulated and minimized as part of our objective.

*Second*, we leverage a regularization based technique from the online learning literature [6] to transform the relaxation of the integer offline optimization problem into a sequence of regularized sub-problems. In particular, the regularization eliminates temporal correlation among decisions across time slots by lifting the precedence constraints coupling successive time slots into the objective function, such that each of the sub-problems can be efficiently and optimally solved in each time slot, using only information at the current time. By solving each sub-problem, VNF instance deployment/removal and flow routing decisions at the time are obtained in polynomial time, constituting part of the feasible (fractional) solution to the relaxed offline problem. Based on the KKT optimality conditions, we are able to show that an upper-bounded overall cost, as compared to the optimal offline solution, can be guaranteed by this (fractional) feasible solution. Moreover, we adapt the regularization framework into a general network flow model to handle the end-to-end delay.

*Third*, we carefully design an online randomized dependent rounding scheme for rounding fractional solutions (on the numbers of VNF instances in each datacenter) to feasible integer solutions of the original problem [7][8]. Our online dependent rounding scheme consists of three modules: 1) A local clustering algorithm that groups datacenters into clusters with small intra-cluster end-to-end delays. One datacenter in each cluster with low resource costs is chosen as the "buffer" datacenter, to deploy VNF instances for absorbing un-served flows due to round-down in the number of VNF deployment in other datacenters. 2) An online dependent rounding algorithm which rounds the fractional numbers of VNF instances to integers while guaranteeing flow routing feasibility. 3) An optimal strategy for intra- and inter-cluster flow redirection based on the rounded solution. Our dependent rounding scheme balances well the minimization among different costs in our objective. Without relying on any future information, it together with the regularization based online algorithm guarantees a good competitive ratio in overall cost as compared to the offline optimum, which is insensitive to the total number of flows nor the number of time slots.

The rest of the paper is organized as follows. We discuss related work in Sec. II and present the problem model and the offline optimization problem in Sec. III. The online algorithm is given in Sec. IV and the dependent rounding scheme is discussed in Sec. V. We present trace-driven evaluation results in Sec. VI and conclude the paper in Sec. VII.

## II. RELATED WORK

Interest on NFV rippled out from a 2012 white paper [1] by telecommunication operators that introduced virtualized network functions running on commodity hardware. Recent IETF drafts presented the use cases of service chains applied across datacenters [2] [9], and highlighted the relation between service chains and NFV. Early efforts on NFV focused on bridging the gap between specialized hardware and network functions [10][11][12], and provided industrial standards for implementing network functions on VMs. Research activities from both industry and academia soon followed suit.

Several NFV management systems have been designed. SIMPLE [13] implements an SDN-based policy enforcement layer for efficient middlebox-specific "traffic steering" in datacenters. Clayman *et al.* [14] design an orchestrator-based architecture for automatic placement of the VNFs. Split/Merge [15] provides system support for achieving efficient, load-balanced elasticity when scaling in and out of virtual middleboxes. OpenNF [3] is a control plane that enables loss-free and order-preserving flow state migration across multiple instances of a VNF, in scenarios where flow packets are distributed across a collection of VNF instances for processing. Stratos [16] is an orchestration layer for efficient and scalable VNF provisioning and scaling via software-defined networking mechanisms. E2 [17] designs an application-agnostic scheduling framework to simplify deployment and scaling of VNFs for packet processing. These systems significantly facilitate the deployment of VNFs, and provide the system bases to support our deployment/scaling algorithm.

A fundamental issue in dynamic VNF provisioning is to optimally place, add and reduce the VNF instances with varying traffic, to provision needed service chains at the minimal cost. There have been a few recent efforts on optimization algorithm design, which mostly deal with one-off placement, ignoring the dynamic nature of an NFV system. VNF-P [18] presents a one-time optimization model for VNF placement, considering hybrid deployment where part of the network service is provided by dedicated hardware and part by VNF instances, and designs a heuristic algorithm. Bari *et al.* [19] study a similar problem, formulate the problem into an integer linear program (ILP), and propose a dynamic programming based heuristic. Mehraghdam *et al.* [20] model a mixed integer quadratically constrained program (MIQCP) to pursue different optimization goals in VNF placement, without giving solution algorithms. Cohen *et al.* [21] investigate one-off VNF placement across different datacenters, to minimize the distance cost between clients and the VNFs that they need and the setup costs of these VNFs, and provide approximation algorithms with rigorous performance analysis. Except [21], all other studies propose heuristic algorithms to solve the respective VNF placement problems, without giving any theoretical performance guarantee. On dynamic VNF scaling, similarly only heuristic approaches are discussed in a few system designs, *e.g.*, Stratos [16] and E2 [17]. Going substantially beyond the existing work, we aim to design an online algorithm for dynamic VNF deployment and scaling across datacenters, and provide solid theoretical guarantee of algorithm efficiency.

## III. PROBLEM MODEL

### A. The NFV System

We consider a network service provider which rents virtual machines (VMs) from cloud datacenters distributed in different geographic locations, deploys virtualized network functions (VNFs) on the VMs, and assembles service chains to serve traffic flows between arbitrary sender and receiver locations. There are in total $M$ types of VNF and $I$ datacenters. Without loss of generality, the system works in a time slotted fashion within a large time span of $1, 2, \ldots, T$. Each time slot represents a decision interval, which is much longer than a typical end-to-end flow delay. Let $[X]$ represent the set $\{1, 2, \ldots, X\}$ throughout the paper, *e.g.*, $[M] = \{1, 2, \ldots, M\}$ is the set of different VNFs.

The VM instances running different VNFs are referred to as *VNF instances*. The flow processing capacities of VNF instances may differ, depending on resource capacities of the respective VMs, including CPU, memory and network I/O. We use $b_{m,i}$ to denote the processing capacity of each instance of VNF $m$ in datacenter $i$, in terms of the maximum flow rate that it can process in each time slot without causing extraordinary processing and queueing delays. Instances of the same VNF are deployed on the same types of VMs in one datacenter, while VM types may differ from one datacenter to another, leading to different processing capacities of instances of the same VNF across different datacenters.

Up to $K$ traffic flows may co-exist at any time in the system, each consisting of data packets from source $s_k$ (location) to destination $z_k$ (location), traversing the same service chain $p_k, k \in [K]$. A service chain $p_k$ contains an ordered sequence of selected VNFs from set $[M]$, *i.e.*, $p_k \subseteq [M]$, and the hops in the service chain can be described by indicators $h_{k,m,m'}, \forall m, m' \in p_k$: $h_{k,m,m'} = 1$, if $m \to m'$ is a hop of $p_k$; and $h'_{k,m,m} = 0$ otherwise. For completeness of formulation, we add a dummy VNF 0 and a dummy VNF $0'$ to the head and the tail of each service chain, respectively, such that $h_{k,0,m} = 1$ if $m$ is the first VNF in the chain and $h_{k,m,0'} = 1$ if $m$ is the last VNF in the chain. A flow can follow different paths from the source to the destination, via different instances of the same VNF along its service chain. Fig. 1 shows two service chains deployed over multiple datacenters, where the flow along service chain 2 can be distributed to three instances of VNF 1 located in 2 different datacenters.

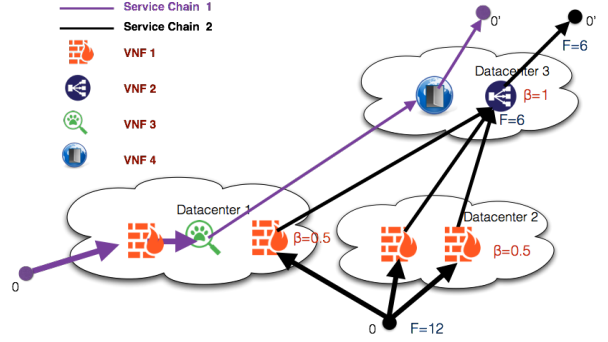

Fig. 1. An example of NFV service chains deployed over geo-distributed datacenters

Let $F_k^{(t)}$ be the rate of flow $k$ at the source, going into instances of the first VNF in its service chain at time $t$. Note that we capture dynamic flow arrivals and departures by allowing the flow rate to be zero if the flow does not exist at the time. Even in the same time slot, the flow rate may change at different hops of the VNF chain: some VNFs perform tunneling gateway functions (*e.g.*, IPSec/SSL VPN and media gateways), converting packets from one format to another, which may increase the packet size for encapsulation or decrease the packet size for decapsulation [22]; some VNFs perform security functions (*e.g.*, firewalls and intrusion detection), dropping packets which violate security policies [23]. We use $\beta_{k,m}$ to denote the average change ratio of flow rates of flow $k$ on VNF $m$, such that the outgoing rate of flow $k$ after traversing an instance of VNF $m$ is on average $\beta_{k,m}$ times the incoming rate. In practice, we can estimate $\beta_{k,m}$ based on the past history and calibrate its value over time. For ease of problem formulation, we define $\bar{\beta}_{k,m}$ as the cumulative rate change ratio of flow $k$ before it goes through VNF $m$, which is the ratio of the overall rate of the flow coming into all instances of VNF $m$ over the initial flow rate $F_k^{(t)}$. $\bar{\beta}_{k,m}$ can be computed using $\beta_{k,m}$'s as follows:

$$\bar{\beta}_{k,m_{k,s}} = 1, \text{ where } m_{k,s} \text{ is the first VNF in service chain } p_k,$$
$$\bar{\beta}_{k,m} = \sum_{m' \in p_k} h_{k,m',m} \bar{\beta}_{k,m'} \beta_{k,m'}, \forall m \in p_k / \{m_{k,s}\}.$$

The total flow rate arriving at all instances of VNF $m$ of flow $k$ at $t$, $\hat{F}_{k,m}^{(t)}$, can be computed based on $F_k^{(t)}$ and $\bar{\beta}_{k,m}$ as

$$\hat{F}_{k,m}^{(t)} = F_k^{(t)} \bar{\beta}_{k,m} \qquad (1)$$

For the example of service chain 2 in Fig. 1, suppose initial rate of the flow going into the service chain is $F_2^{(1)} = 12$; then we have $\hat{F}_{2,1}^{(1)} = F_2^{(1)} = 12$, $\hat{F}_{2,2}^{(1)} = 6$ (since $\beta_{2,1} = 0.5$) and $\hat{F}_{2,0'}^{(1)} = 6$ (since $\beta_{2,2} = 1$).

## B. Decision Variables

We seek to make the following decisions on VNFs deployment and flow routing at each time slot $t \in [T]$: (i) $q_{m,i}^{(t)}$, the number of instances of VNF $m$ to deploy in datacenter $i$, $\forall m \in [M], i \in [I]$. An instance of a VNF can be shared by multiple flows whose service chain includes this VNF. (ii) $x_{k,m,i,m',i'}^{(t)}$, the overall egress rate of flow $k$ from instances of VNF $m$ at datacenter $i$ to instances of VNF $m'$ at datacenter $i'$, $\forall k \in [K], m, m' \in [M], i, i' \in [I]$. For ease of optimization formulation, we also introduce $y_{k,m,i}^{(t)}$ to denote the total ingress rate of flow $k$ to instances of VNF $m$ at datacenter $i$, i.e., $y_{k,m,i}^{(t)} = \sum_{i' \in [I]/\{i\}} \sum_{m' \in [M]} h_{k,m',m} x_{k,m',i',m,i}^{(t)}$. Here $q_{m,i}^{(t)}$'s are non-negative integers, while $x_{k,m,i,m',i'}^{(t)}$'s and $y_{k,m,i}^{(t)}$'s are non-negative real numbers. In the example in Fig. 1, $q_{1,1}^{(1)} = 2$, $q_{1,2}^{(1)} = 2$ and $q_{2,3}^{(1)} = 1$; for service chain 2, supposing the incoming flow is evenly distributed to the three instances of VNF 1 in datacenters 1 and 2, then $x_{2,1,1,2,3}^{(1)} = 2$, $x_{2,1,2,2,3}^{(1)} = 4$ and $y_{2,2,3}^{(1)} = 6$.

## C. Cost Structure

We aim to minimize the overall operational cost of the network service provider during $[T]$, with the following costs considered.

1) *VNF Running Cost.* The network service provider pays the cloud operator for renting VMs to run the VNF instances. Let $c_{m,i}^{(t)}$ be the cost of operating an instance of VNF $m$ in datacenter $i$ in time slot $t$. The overall cost for VM rentals in all datacenters in $t$ is:

$$C_R^{(t)} = \sum_{i \in [I]} \sum_{m \in [M]} c_{m,i}^{(t)} q_{m,i}^{(t)} \quad (2)$$

2) *VNF Deployment Cost.* Launching a new VNF instance commonly involves transferring a VM image containing the network function to the hosting datacenter/server, and booting and attaching the image to device. The deployment cost is typically considered on the order of the cost to run a server for a number of seconds or minutes [24]. Let $\delta_{m,i}$ denote the cost for deploying an instance of VNF $m$ in datacenter $i$. Let $\rho_{m,i}^{(t)}$ denote the number of new instances of VNF $m$ to be deployed in datacenter $i$ in $t$, which can be calculated as

$$\rho_{m,i}^{(t)} = \max\{0, q_{m,i}^{(t)} - q_{m,i}^{(t-1)}\} \quad (3)$$

The total deployment cost for all new VNF instances in all datacenters in $t$ is hence computed as

$$C_D^{(t)} = \sum_{i \in [I]} \sum_{m \in [M]} \delta_{m,i} \rho_{m,i}^{(t)} \quad (4)$$

3) *Flow Transfer Cost.* To transfer flows into and out of a datacenter, a bandwidth charge may be incurred (*e.g.*, [25]). Let $d_i^{in}$ and $d_i^{out}$ respectively be the costs of sending a unit of a flow into and out of datacenter $i$ in a time slot. The overall bandwidth cost of all flows across all datacenters in $t$ is:

$$C_T^{(t)} = \sum_{k \in [K]} \sum_{m' \in [M]} \sum_{i' \in [I]/\{i\}} \sum_{m \in [M]} \sum_{i \in [I]} h_{k,m',m} d_i^{in} x_{k,m',i',m,i}^{(t)}$$
$$+ \sum_{k \in [K]} \sum_{m \in [M]} \sum_{i \in [I]} \sum_{m' \in [M]} \sum_{i' \in [I]/\{i\}} h_{k,m,m'} d_i^{out} x_{k,m,i,m',i'}^{(t)}$$

which is equivalent to

$$C_T^{(t)} = \sum_{k \in [K]} \sum_{m \in [M]} \sum_{i \in [I]} (d_i^{in} + d_i^{out} \beta_{k,m}) y_{k,m,i}^{(t)}$$
$$- \sum_{k \in [K]} \sum_{m \in [M]} \sum_{m' \in [M]} \sum_{i \in [I]} h_{k,m,m'} (d_i^{in} + d_i^{out}) x_{k,m,i,m',i}^{(t)}$$

where the deduction is to remove flows between VNF instances in the same datacenter from the cost computation.

4) *End-to-end Flow Delay.* As an important performance indicator, the end-to-end delay of each flow to pass through its service chain should be minimized, which is mainly decided by locations of the VNF instances that it traverses, *i.e.*, delays on the hops. Let $l_{i,j}$ denote the delay between node $i$ and node $j$ in the set of all datacenters, flow sources, and flow destinations, such that $l_{i,j} = l_{j,i}$ and $l_{i,i} = 0$. We use a generalized $\alpha-$relaxed triangle inequality [26] to describe delays between nodes:

$$|l_{a,b} - l_{b,c}| \leq \alpha l_{a,c}, \forall a, b, c \in [I] \cup_{k \in [K]} \{s_k, z_k\} \quad (5)$$

If $\alpha \leq 1$, (5) becomes the normal triangle inequality, $l_{a,b} \leq l_{a,c} + l_{b,c}$; $\alpha > 1$ implies violation of the normal triangle inequality, which is common for the Internet delay space.

The average end-to-end delay of a flow $k$ can be computed by summing up the following: (i) The average delay from source $s_k$ to datacenters hosting the first VNF of service chain $p_k$, $\sum_{m \in [M]} h_{k,0,m}(\frac{\sum_{i \in [I]} l_{s_k,i} y_{k,m,i}^{(t)}}{F_k^{(t)}})$. The summation over all datacenters represents that the flow can be dispatched to different instances of the first VNF $m$ ($h_{k,0,m} = 1$) in different datacenters, and dividing the sum by $F_k^{(t)}$ removes the impact of flow rate, leaving only delay in the result. (ii) The average delay in each hop $m \to m'$ of service chain $p_k$ ($h_{k,m,m'} = 1$), $h_{k,m,m'} \frac{\sum_{i \in [I]} \sum_{i' \in [I]} l_{i,i'} x_{k,m,i,m',i'}^{(t)}}{\beta_{k,m} \bar{\beta}_{k,m} F_k^{(t)}}$. The summation over all datacenter pairs models that both VNF $m$ and VNF $m'$ can be deployed in multiple datacenters, and dividing the sum by the overall flow rate at this hop, $\beta_{k,m} \bar{\beta}_{k,m} F_k^{(t)}$, removes the influence of flow rates in delay calculation. (iii) The average delay from the datacenters hosting instances of the last VNF of service chain $p_k$ to destination $d_k$, $\sum_{m \in [M]} h_{k,m,0'}(\frac{\sum_{i \in [I]} l_{i,z_k} \beta_{k,m} y_{k,m,i}^{(t)}}{\beta_{k,m} \bar{\beta}_{k,m} F_k^{(t)}})$. Hence the average end-to-end delay of flow $k$ in $t$ is:

$$\sum_{m \in [M]} \sum_{i \in [I]} \frac{h_{k,0,m} l_{s_k,i}}{F_k^{(t)}} y_{k,m,i}^{(t)} + \sum_{m \in [M]} \sum_{i \in [I]} \frac{h_{k,m,0} l_{i,z_k}}{\bar{\beta}_{k,m} F_k^{(t)}} y_{k,m,i}^{(t)}$$
$$+ \sum_{m \in [M]} \sum_{i \in [I]} \sum_{m' \in [M]} \sum_{i' \in [I]} \frac{h_{k,m,m'} l_{i,i'}}{\beta_{k,m} \bar{\beta}_{k,m} F_k^{(t)}} x_{k,m,i,m',i'}^{(t)}$$

By multiplying the average end-to-end delay of each flow $k$ with a weight parameter $a_k^{(t)}$, we convert the delay to a delay cost, to be aggregated with other costs for minimization in our optimal decision making. The overall delay cost of all flows in time slot $t$ is:

$$C_E^{(t)} = \sum_{k \in [K]} \sum_{m \in [M]} \sum_{i \in [I]} \xi_{k,m,i}^{(t)} y_{k,m,i}^{(t)}$$
$$+ \sum_{k \in [K]} \sum_{m \in [M]} \sum_{i \in [I]} \sum_{m' \in [M]} \sum_{i' \in [I]} \omega_{k,m,i,m',i'}^{(t)} x_{k,m,i,m',i'}^{(t)} \quad (6)$$

where $\xi_{k,m,i}^{(t)} = \frac{a_k^{(t)} h_{k,0,m} l_{s_k,i}}{F_k^{(t)}} + \frac{a_k^{(t)} h_{k,m,0'} l_{i,z_k}}{\bar{\beta}_{k,m} F_k^{(t)}}$ and

$$\omega_{k,m,i,m',i'}^{(t)} = \frac{a_k^{(t)} h_{k,m,m'} l_{i,i'}}{\beta_{k,m} \bar{\beta}_{k,m} F_k^{(t)}}.$$

Such a conversion of delay to a delay cost, in order to sum it up in the total cost for minimization, can be interpreted from the angle of multi-objective optimization [27]: the classic technique to treat an optimization problem where we have more than one objectives (*e.g.*, minimize operating cost and minimize delay as in our case), *i.e.*, a multi-objective optimization problem, is to convert it into a single objective optimization problem, where the new objective is a weighted sum of the original multiple objectives.

### D. The Offline Cost Minimization Problem

Assuming full knowledge of the system in $[T]$, we can formulate an offline VNF deployment and flow routing problem in (7), for overall cost minimization. Important notation is listed in Table I for ease of reference.

$$\mathbf{P}: \quad \text{minimize} \sum_{t \in [T]} C_R^{(t)} + C_D^{(t)} + C_T^{(t)} + C_E^{(t)} \quad (7)$$

subject to:

$$\sum_{k \in [K]} y_{k,m,i}^{(t)} \leq q_{m,i}^{(t)} b_{m,i}, \forall t \in [T], i \in [I], m \in [M] \quad (8a)$$

$$\rho_{m,i}^{(t)} \geq q_{m,i}^{(t)} - q_{m,i}^{(t-1)}, \forall t \in [T], i \in [I], m \in [M] \quad (8b)$$

$$\sum_{i \in [I]} y_{k,m,i}^{(t)} = \hat{F}_{k,m}^{(t)}, \forall t \in [T], m \in [M], k \in [K] \quad (8c)$$

$$y_{k,m,i}^{(t)} = \sum_{i' \in [I]} \sum_{m' \in [M]} h_{k,m',m} x_{k,m',i',m,i}^{(t)} \quad (8d)$$

$$\forall t \in [T], m \in [M]/\{m_{k,s}\}, i \in [I], k \in [K]$$

$$\beta_{k,m} y_{k,m,i}^{(t)} = \sum_{m' \in [M]} \sum_{i' \in [I]} h_{k,m,m'} x_{k,m,i,m',i'}^{(t)} \quad (8e)$$

$$\forall t \in [T], m \in [M]/\{m_{k,z}\}, i \in [I], k \in [K]$$

$$x_{k,m,i,m',i'}^{(t)} \geq 0, \quad (8f)$$
$$\forall t \in [T], m, m' \in [M], i, i' \in [I], k \in [K]$$

$$y_{k,m,i}^{(t)} \geq 0, \quad \forall t \in [T], k \in [K], m \in [M], i \in [I] \quad (8g)$$

$$q_{m,i}^{(0)} = 0, \quad \forall m \in [M], i \in [I], \quad (8h)$$

$$q_{m,i}^{(t)} \in \{0, 1, 2, \ldots\}, \quad \forall t \in [T], m \in [M], i \in [I] \quad (8i)$$

$$\rho_{m,i}^{(t)} \in \{0, 1, 2, \ldots\}, \quad \forall t \in [T], m \in [M], i \in [I] \quad (8j)$$

Here $m_{k,s}$ ($m_{k,z}$) denotes the first (last) VNF in flow $k$'s service chain. Constraint (8a) guarantees that at each time, the total incoming flow rate to instances of VNF $m$ in datacenter $i$ does not exceed the processing capacity of the deployed instances. Constraints (8b) and (8j) are derived from the definition of auxiliary variable $\rho_{m,i}^{(t)}$ in (3). (8c) shows that the total incoming rate of flow $k$ to VNF $m$ in all datacenters should equal the (changed) aggregate rate of the flow at this hop, where $\hat{F}_{k,m}^{(t)}$ is decided by the initial flow rate and the change ratios along the service chain as in Eqn. (1). (8d) and (8e) guarantee flow conservation (considering flow rate change ratios $\beta_{k,m}$) for each flow at instances of each VNF in each datacenter, by connecting ingress and egress flows.

The offline cost minimization problem (7) is a mixed integer linear program (MILP). Even with complete information of new flow arrivals and flow rates in the entire span $[T]$, solving such an MILP is non-trivial. The problem can be classified as the multiple-knapsack problem (MKP), which is known to be NP-hard, and more strongly, has no fully polynomial time approximation schemes unless P=NP [28].

TABLE I
KEY NOTATION

| | |
|---|---|
| $I$ | # of datacenters |
| $M$ | # of VNF types |
| $T$ | # of time slots |
| $K$ | # of flows |
| $s_k, z_k$ | source / destination of flow $k$ |
| $F_k^{(t)}$ | flow rate of flow $k$ at time $t$ |
| $\hat{F}_{k,m}$ | total rate of flow $k$ through VNF $m$ |
| $p_k$ | service chain that flow $k$ follows |
| $a_k^{(t)}$ | delay-to-cost conversion coefficient for flow $k$ at $t$ |
| $h_{k,m,m'}$ | indicator of whether $m \to m'$ is a hop of $p_k$ |
| $l_{i,j}$ | delay from $i$ to $j$ |
| $\beta_{k,m}$ | rate change ratio of flow $k$ through VNF $m$ |
| $\bar{\beta}_{k,m}$ | accumulative rate change ratio of flow $k$ before going through VNF $m$ |
| $y_{k,m,i}^{(t)}$ | total ingress flow rate of flow $k$ to VNF $m$ in datacenter $i$ at $t$ |
| $x_{k,m,i,m',i'}^{(t)}$ | rate of flow $k$ from VNF $m$ in datacenter $i$ to VNF $m'$ in datacenter $i'$ at $t$ |
| $b_{m,i}$ | processing capacity per instance of VNF $m$ in datacenter $i$ |
| $q_{m,i}^{(t)}$ | # of instances of VNF $m$ deployed in datacenter $i$ in $t$ |
| $\rho_{m,i}^{(t)}$ | # of new instances of VNF $m$ deployed in datacenter $i$ in $t$ |
| $\delta_{m,i}$ | deployment cost for each new instance of VNF $m$ in datacenter $i$ |
| $c_{m,i}^{(t)}$ | operational cost for running an instance of VNF $m$ in datacenter $i$ in $t$ |
| $d_i^{in}, d_i^{out}$ | flow transfer cost for transferring a unit of flow in one time slot in/out of datacenter $i$ |

In an online setting, all parameters, variables and constraints related to time slot $t$ are revealed upon the arrival of the corresponding time. Due to constraints (8b), decisions of one time slot are coupled with those in another. We seek to design an efficient online algorithm that produces VNF deployment and flow routing decisions based only on the current flow information and past history, while achieving a good competitive ratio, computed by dividing the overall cost achieved with our online algorithm by the offline minimum cost derived by solving (7) exactly.

Our design of the online algorithm is divided into two steps. First, we relax the integrality constraints (8i) and (8j) in (7), obtain the fractional offline optimization problem $P_f$ as follows, and apply a novel regularization method to design an online algorithm for solving this relaxed linear program (LP) in Sec. IV. We then design a dependent rounding algorithm to round the fractional solutions to feasible solutions of MILP (7) in Sec. V.

$$\mathbf{P_f}: \quad \text{minimize} \sum_{t \in [T]} C_R^{(t)} + C_D^{(t)} + C_T^{(t)} + C_E^{(t)} \quad (9)$$

subject to:
constraints (8a) to (8h)

$$q_{m,i}^{(t)} \geq 0, \quad \forall t \in [T], m \in [M], i \in [I] \quad (9i)$$

$$\rho_{m,i}^{(t)} \geq 0, \quad \forall t \in [T], m \in [M], i \in [I]$$

**Algorithm 1:** An Online Regularization-based Fractional Algorithm - *ORFA*

    **Input:** $K, M, I, \beta, \delta, \mathbf{h}, \mathbf{b}, \mathbf{s}, \mathbf{l}, \mathbf{d}^{in}, \mathbf{d}^{out}, \epsilon$
    **Output:** $\mathbf{q}, \mathbf{x}, \mathbf{y}$
1   Initialization: $\mathbf{q} = \mathbf{0}, \mathbf{x} = \mathbf{0}, \mathbf{y} = \mathbf{0}$;
2   **for** *each time slot* $t \in [T]$ **do**
3      observe values of $F_k^{(t)}, a_k^{(t)}, c_{m,i}^{(t)}$,
        $\forall k \in [K], m \in [M], i \in [I]$;
4      compute $\hat{F}_k^{(t)}, \forall k \in [K]$, according to (1);
5      use the interior point method to solve $\tilde{P}_f^{(t)}$ in (11);
6      **return** optimal solutions $\mathbf{q}^{(t)}, \mathbf{x}^{(t)}, \mathbf{y}^{(t)}$;
7 **end**

## IV. THE FRACTIONAL ONLINE ALGORITHM VIA REGULARIZATION

**Regularization.** The key idea of our online algorithm to solve $P_f$ is to lift constraint (8b) to the objective function using a smooth convex function. By doing this, decisions in time slots $t-1$ and $t$ can be decoupled and the resulting relaxed offline problem, denoted by $\tilde{P}_f$, can be readily decomposed into a set of sub-problems, each to be efficiently solved in one time slot.

To lift (8b) into the objective function in (7), we observe that (8b) originates from the definition of variable $\rho_{m,i}^{(t)}$ in (3), the number of new instances of VNF $m$ to be deployed in datacenter $i$ in $t$. $\rho_{m,i}^{(t)}$ is involved in the deployment cost $C_D^{(t)}$ as defined in (4), which is the second term of the objective function in (7). We substitute $\rho_{m,i}^{(t)}$ in (7) by a function of $q_{m,i}^{(t-1)}$ and $q_{m,i}^{(t)}$ via the regularization technique from the online learning literature [29], removing constraint (8b). In particular, we use the following relative entropy function, a commonly adopted convex regularizer, as the basis of the function to approximate $\rho_{m,i}^{(t)}$ in (3):

$$\Delta(q_{m,i}^{(t)} \| q_{m,i}^{(t-1)}) = q_{m,i}^{(t)} \ln \frac{q_{m,i}^{(t)}}{q_{m,i}^{(t-1)}} + q_{m,i}^{(t-1)} - q_{m,i}^{(t)} \quad (10)$$

This relative entropy function is obtained by summing the relative entropy ($q_{m,i}^{(t)} \ln \frac{q_{m,i}^{(t)}}{q_{m,i}^{(t-1)}}$) and a linear term representing the movement cost ($q_{m,i}^{(t-1)} - q_{m,i}^{(t)}$). Such a function, proven convex, is a widely adopted regularizer in online learning problems that involve $l_1$-norm constraints [30][29], such as in our case (constraint (8b)). Further, to approximate $\rho_{m,i}^{(t)}$ in (3), we add a constant term $\frac{\epsilon}{MI}$ to both $q_{m,i}^{(t)}$ and $q_{m,i}^{(t-1)}$ in the relative entropy term in (10), where $\epsilon$ is a small positive constant, to ensure that the fraction is still valid when the number of VNF instances deployed in $t-1$ is zero, *i.e.*, $q_{m,i}^{(t-1)} = 0$. We also define a parameter $\eta = \ln(1 + \frac{MI}{\epsilon})$ and multiply the improved relative entropy function by $\frac{1}{\eta}$, which is related to our competitive ratio, to normalize the deployment cost by regularization.

We replace $\rho_{m,i}^{(t)}$ in (7) by the constructed function, and obtain $\tilde{P}_f$ as follows. Here we only present the subproblem at $t$, $\tilde{P}_f^{(t)}$, that $\tilde{P}_f$ is decomposed into, $\forall t \in [T]$. $\tilde{P}_f$ can be readily obtained by adding a summation over $t \in [T]$ in the objective function, *i.e.*, $\tilde{P}_f = \sum_{t \in [T]} \tilde{P}_f^{(t)}$, and repeating each constraint for each $t \in [T]$.

$$\tilde{\mathbf{P}}_\mathbf{f}^{(\mathbf{t})}: \quad \text{minimize } C_R^{(t)} + C_T^{(t)} + C_E^{(t)} \quad (11)$$

$$+ \sum_{m \in [M]} \sum_{i \in [I]} \frac{\delta_{m,i}}{\eta} \left( \left( q_{m,i}^{(t)} + \frac{\epsilon}{MI} \right) \ln \frac{q_{m,i}^{(t)} + \frac{\epsilon}{MI}}{q_{m,i}^{(t-1)} + \frac{\epsilon}{MI}} + q_{m,i}^{(t-1)} - q_{m,i}^{(t)} \right)$$

subject to: constraints (8a)(8c)-(8h), (9i), for $t$

**Online Algorithm.** Our online algorithm to derive a solution to $P_f$ in (9), given in Alg. 1, runs by solving subproblem $\tilde{P}_f^{(t)}$ in (11) in each time slot $t$. Note that $q_{m,i}^{(t-1)}, \forall i \in [I], m \in [M]$, have been obtained when solving $\tilde{P}_f^{(t-1)}$ in time $t-1$, and their values are given as input to $\tilde{P}_f^{(t)}$ at $t$. Since $\tilde{P}_f^{(t)}$ is a convex optimization problem with linear constraints, it can be optimally solved in polynomial time, *e.g.*, using the interior point method [27]. It is obvious that the optimal solution of $\tilde{P}_f^{(t)}, \forall t \in [T]$, constitute a feasible solution to $P_f$ (values of $\rho_{m,i}^{(t)}$'s can be easily set based on (3)), as given in the following theorem.

**Theorem 1.** *The online fractional algorithm ORFA produces a feasible solution of $P_f$ in polynomial time.*

**Competitive Ratio.** $q_{m,i}^{(t)}$'s derived by *ORFA* are potentially fractional. We next show that the fractional solution can achieve a good competitive ratio in overall cost. In Sec. V, we will round the fractional solutions to integers, as well as show the final competitive ratio based on the ratio here plus the efficiency loss due to rounding.

**Theorem 2.** *The overall cost in (7) that the solution derived by ORFA achieves is at most $\log(1 + MI/\epsilon) + 1 + 1/\phi$ times the offline minimum overall cost derived by solving $P$ in (7) exactly, where $\phi = \min_{t \in [T], i \in [I], m \in [M]} \{q_{m,i}^{(t)} : q_{m,i}^{(t)} > 0\}$ is the minimum positive number of instances deployed for any VNF, at any time slot and in any datacenter.*

The detailed proof is given in Appendix A.

We next give a bound on the value of the objective function achieved by any integer solution of (11), which will be used in our analysis of the final competitive ratio in the next section.

**Theorem 3.** *The objective value of $\tilde{P}_f$ in (11) achieved by the best integral solution of (11), denoted by $\tilde{P}_I$, is at most $\log(1+MI/\epsilon)+2$ times the offline minimum overall cost.*

Theorem 3 follows immediately Theorem 2, since the smallest integer value of $\phi$ in the ratio given in Theorem 2 is 1, denoting the minimum positive integral number of instances deployed for any VNF in any datacenter at any time.

## V. AN ONLINE DEPENDENT ROUNDING SCHEME

### A. The Dependent Rounding Scheme

The ORFA algorithm in Sec. IV computes a fractional solution $(\mathbf{q}^{(t)}, \mathbf{x}^{(t)}, \mathbf{y}^{(t)}, \rho^{(t)})$, for the optimization problem in (9). However, the number of instances of each VNF to deploy should be integral, as captured by (8i) and (8j) in the offline problem. We need to round the fractional solution $\mathbf{q}^{(t)}$ to an integer solution $\bar{\mathbf{q}}^{(t)}$. Due to constraints (8a) - (8e), modifying $\mathbf{q}^{(t)}$ requires correspondingly modifying $(\mathbf{x}^{(t)}, \mathbf{y}^{(t)}, \rho^{(t)})$ to maintain solution feasibility. Our goal in this section is to compute a rounded solution $(\bar{\mathbf{q}}^{(t)}, \bar{\mathbf{x}}^{(t)}, \bar{\mathbf{y}}^{(t)}, \bar{\rho}^{(t)})$ from $(\mathbf{q}^{(t)}, \mathbf{x}^{(t)}, \mathbf{y}^{(t)}, \rho^{(t)})$ such that the rounded solution is feasible to (7).

We first note that a straightforward independent rounding scheme, where each variable is rounded up or down independently of other variables, may violate feasibility. For example, there is a chance that $q_{m,i}^{(t)}, \forall m \in [M], i \in [I]$, are rounded down, making it impossible to obtain a feasible routing solution. We therefore design a dependent rounding scheme that can exploit the inherent dependence of the variables in capacity constraint (8a) and flow conservation constraints (8d) and (8e). The key idea is that rounded-down VNF instance numbers will be compensated by rounded-up VNF instance numbers, as well as extra VNF instances deployed, guaranteeing a feasible $\bar{\mathbf{y}}^{(t)}$ and a feasible $\bar{\mathbf{x}}^{(t)}$ that satisfy (8a), (8d) and (8e).

Our dependent rounding scheme contains four modules: (i) a local clustering algorithm that separates the geo-distributed datacenter network into several clusters, each with at least two datacenters; (ii) an initialization module for weighted dependent rounding, that constructs a weighted star graph in each time slot; (iii) a weighted dependent rounding algorithm based on the clusters and start graph above to compute the rounded solution of $\bar{\mathbf{q}}^{(t)}$ and $\bar{\rho}^{(t)}$; (iv) a flow direction algorithm to find the feasible solution of $\bar{\mathbf{y}}^{(t)}$ and $\bar{\mathbf{x}}^{(t)}$, which will provide $(\bar{\mathbf{q}}^{(t)}, \bar{\mathbf{x}}^{(t)}, \bar{\mathbf{y}}^{(t)}, \bar{\rho}^{(t)})$ as the final feasible solution to (7).

*1) Datacenter Clustering:* Geo-distributed datacenters in practice often form natural clusters based on location proximity. Amazon EC2 datacenters [31] and Google datacenters [32] can be separated into 3 main clusters by continents: America, Europe and Asia datacenter groups. Delays between datacenters within the same cluster (intra-cluster delays) are substantially smaller than those between datacenters across different clusters (inter-cluster delays). We design a local clustering algorithm, as given in Alg. 2, to group datacenters in the system into clusters with intra-cluster delays no larger than $R$. Here $R$ is defined to be the median of all the inter-datacenter delays (line 2). In Alg. 2, we construct the clusters by repeatedly merging existing clusters whose inter-cluster delays are no larger than $R$. Then we identify the isolated points (clusters including only one datacenter) and merge them to nearest clusters. The clustering algorithm is carried out once at the beginning of the system. By carrying out dependent rounding on top of the cluster structure, we will be able to bound the delay cost part in the overall cost in our competitive analysis, where the bound is related to $R$.

---

**Algorithm 2:** The Local Clustering Algorithm

**Input:** $l_{i,i'}, \forall i, i' \in [I]$
**Output:** clusters of datacenters $\odot$
1 Treat each datacenter $i$ as a singleton cluster;
2 $R = median_{i,i' \in [I]} l_{i,i'}$;
3 **while** *there are more clusters to merge* **do**
4    **for** *Any two clusters $\odot_u \neq \odot_v$* **do**
5      **if** $l_{i,j} \leq R \ \forall i \in \odot_u, j \in \odot_v$ **then**
6        Merge $\odot_u$ and $\odot_v$ into a new cluster $\odot_{u'}$.
7      **end**
8    **end**
9 **end**
10 **for** *any cluster $\odot_u$ with only one datacenter $i$* **do**
11    Find the cluster $\odot_v$ with minimal value of $max_{j \in \odot_v} l_{i,j}$, merge $\odot_u$ into $\odot_v$.
12 **end**

---

*2) Star Graph Construction:* In each time slot, we construct a star graph for each datacenter cluster (Alg. 3). We identify a buffer datacenter, $j_m^{(t)*}$ for each type of VNF $m$ in each datacenter cluster, which has the smallest VNF running cost per unit flow processing capacity (line 4). We may deploy extra instances of VNF $m$ in such a buffer datacenter to absorb unserved flow due to round-down of VNF instance numbers in other non-buffer datacenters in the cluster. The buffer datacenter is the centre of the star graph for the cluster. Let $A_{m,j_m^{(t)*}}$ denote the non-buffer datacenters in cluster $\odot_{m,j_m^{(t)*}}$, $B_{m,j_m^{(t)*}}$ denote the buffer datacenter in $\odot_{m,j_m^{(t)*}}$, and $\Upsilon_{m,j_m^{(t)*}}$ denote the set of fractional edges in $\odot_{m,j_m^{(t)*}}$ with $0 < p_{m,i}^{(t)} < 1$ (line 6). For each datacenter in each cluster except the buffer (line 7), if the computed instance number of VNF $m$ by the online algorithm ORFA is already an integer, we ignore this datacenter in the rounding scheme and remove it from $\Upsilon$ (lines 8-9); otherwise, we add an edge connecting the datacenter to the buffer datacenter, and associate a probability $p_{m,i}^{(t)}$ and a weight with the edge $w_{m,i}^{(t)}$ (lines 13-14). $p_{m,i}^{(t)}$ is computed by (12) and will be used for rounding. $w_{m,i}^{(t)}$ is computed by (13), and will be used for determining the number of instances of VNF $m$ to be deployed in the buffer datacenter. After running this initialization module, the data center graph is dissected into disjoint stars, as illustrated in Fig. 2.

*3) Weighted Dependent Rounding:* We are now ready to present the Online Weighted Dependent Rounding (OWDR) algorithm in Alg. 4, which computes a feasible integer solution of $\bar{q}_{m,i}^{(t)}$ in each time slot $t$. For each type of VNF $m$, OWDR runs a series of rounding iterations for each cluster (line 1). For each cluster $\odot_{m,j_m^{(t)*}}$,

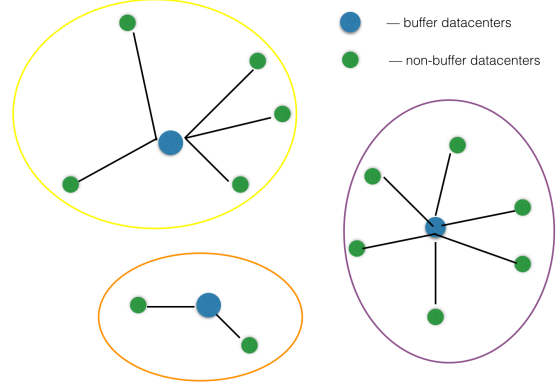

Fig. 2. Datacenter graph with disjoint stars

---

**Algorithm 3:** The Initialization Module for Weighted Dependent Rounding in $t$, *INIT*

**Input:** $K, M, I, \mathbf{q}^{(t)}, \mathbf{c}, \mathbf{b}$
**Output:** $\mathbf{w}^{(t)}, \mathbf{p}^{(t)}, \mathbf{A}, \mathbf{B}, \boldsymbol{\Upsilon}, G(V_m^{(t)}, E_m^{(t)}), \forall m \in [M]$
1 **for** *each type of VNF $m \in [M]$* **do**
2    Set $V_m^{(t)} = [I], E_m^{(t)} = \emptyset$;
3    **for** *each datacenter cluster* **do**
4      Decide datacenter $j_m^{(t)*}$ with minimal $\frac{c_{m,i}^{(t)}}{b_{m,i}}$; (if there is a tie, choose any one from the tied candidates)
5      Label this datacenter cluster as $\odot_{m,j_m^{(t)*}}$;
6      Set $A_{m,j_m^{(t)*}} = \odot_{m,j_m^{(t)*}}/\{j_m^{(t)*}\}$, $B_{m,j_m^{(t)*}} = \{j_m^{(t)*}\}$, $\Upsilon_{m,j_m^{(t)*}} = \{(i, j_m^{(t)*})\} \quad \forall i \in A_{m,j_m^{(t)*}}$;
7      **for** *each datacenter $i \in A_{m,j_m^{(t)*}}$* **do**
8        **if** $q_{m,i}^{(t)} = \lfloor q_{m,i}^{(t)} \rfloor$ **then**
9          $\Upsilon_{m,j_m^{(t)*}} = \Upsilon_{m,j_m^{(t)*}}/\{(i, j_m^{(t)*})\}$;
10        **end**
11        **else**
12          Add the edge $(i, j_m^{(t)*})$ to $E_m^{(t)}$;
13          Associate a probability coefficient
$$p_{m,i}^{(t)} = \lfloor q_{m,i}^{(t)} \rfloor + 1 - q_{m,i}^{(t)} \quad (12)$$
with the edge $(i, j_m^{(t)*})$;
14          Associate a weight coefficient
$$w_{m,i}^{(t)} = (q_{m,i}^{(t)} - \lfloor q_{m,i}^{(t)} \rfloor)\frac{b_{m,i}}{b_{m,j_m^{(t)*}}} \quad (13)$$
with the edge $(i, j_m^{(t)*})$;
15        **end**
16      **end**
17    **end**
18 **end**

---

we find a maximum path in its star topology (of length at most 2) that will be the path from a datacenter in $A_{m,j_m^{(t)*}}$ to the buffer datacenter $j_m^{(t)*}$ to another datacenter in $A_{m,j_m^{(t)*}}$. Then the two edges are placed into two distinct matchings (line 4). If only one floating edge remains (this can happen only in the last iteration), we choose this edge as the maximal path with length 1. During each iteration, we let the probability of one of these two edges, $p_{m,i_1}^{(t)}$ or $p_{m,i_2}^{(t)}$ round to 0 or 1, decided by the coupled coefficient $\kappa_1$ and $\kappa_2$, which will decrease the number of elements in $\Upsilon_{m,j_m^{(t)*}}$ by 1. Finally, the

**Algorithm 4:** The Online Weighed Dependent Rounding Algorithm (OWDR) in $t$

**Input:** $\mathbf{q}^{(t)}$, star graphs $G(V_m^{(t)}, E_m^{(t)}), \forall m \in [M]$, $\mathbf{p}^{(t)}, \mathbf{w}^{(t)}, \bar{\mathbf{q}}^{(t-1)}$
**Output:** $\bar{\mathbf{q}}^{(t)}, \bar{\rho}^{(t)}$

1 **for** *each type of VNF* $m \in [M]$ **do**
2  **for** *each datacenter cluster* $\odot_{m,j_m^{(t)*}}$ **do**
3   **while** $\Upsilon_{m,j_m^{(t)*}} \neq \emptyset$ **do**
4    Run the depth-first-search (DFS) algorithm to obtain maximal path $\mathbb{P}$ in the star graph $(A, B, \Upsilon)$ and partition the edge set of $\mathbb{P}$ into two matchings $\Psi_1$ and $\Psi_2$ ;
6
5    Define
$$\kappa_1 = \min\{\gamma > 0 | (\exists (i_1, j_m^{(t)*}) \in \Psi_1 : p_{m,i_1}^{(t)} + \gamma = 1)$$
$$\vee (\exists (i_2, j_m^{(t)*}) \in \Psi_2 : p_{m,i_2}^{(t)} - \frac{w_{m,i_1}^{(t)}}{w_{m,i_2}^{(t)}} \gamma = 0)\}$$
$$\kappa_2 = \min\{\gamma > 0 | \exists i_1, j_m^{(t)*}) \in \Psi_1 : p_{m,i_1}^{(t)} - \gamma = 0)$$
$$\vee (\exists i_2, j_m^{(t)*}) \in \Psi_2 : p_{m,i_2}^{(t)} + \frac{w_{m,i_2}^{(t)}}{w_{m,i_1}^{(t)}} \gamma = 1)\}$$

7    With probability $\frac{\kappa_2}{\kappa_1+\kappa_2}$, set
$$p_{m,i_1}^{(t)} = p_{m,i_1}^{(t)} + \kappa_1, \quad \forall (i_1, j_m^{(t)*}) \in \Psi_1$$
$$p_{m,i_2}^{(t)} = p_{m,i_2}^{(t)} - \kappa_1 \frac{w_{m,i_1}^{(t)}}{w_{m,i_2}^{(t)}}, \quad \forall (i_2, j_m^{(t)*}) \in \Psi_2$$

     Remove $(i_1, j_m^{(t)*})$ or $(i_2, j_m^{(t)*})$ from $\Upsilon_{m,j_m^{(t)*}}$ if $p_{m,i_1}^{(t)} \in \{0,1\}$ or $p_{m,i_2}^{(t)} \in \{0,1\}$
8    With probability $\frac{\kappa_1}{\kappa_1+\kappa_2}$, set
$$p_{m,i_1}^{(t)} = p_{m,i_1}^{(t)} - \kappa_2, \quad \forall (i_1, j_m^{(t)*}) \in \Psi_1$$
$$p_{m,i_2}^{(t)} = p_{m,i_2}^{(t)} + \kappa_2 \frac{w_{m,i_1}^{(t)}}{w_{m,i_2}^{(t)}}, \quad \forall (i_2, j_m^{(t)*}) \in \Psi_2$$

     Remove $(i_1, j_m^{(t)*})$ or $(i_2, j_m^{(t)*})$ from $\Upsilon_{m,j_m^{(t)*}}$ if $p_{m,i_1}^{(t)} \in \{0,1\}$ or $p_{m,i_2}^{(t)} \in \{0,1\}$
9   **end**
10  **end**
11  **for** *each datacenter* $i \in A$ **do**
12
$$\bar{q}_{m,i}^{(t)} = \lfloor q_{m,i}^{(t)} \rfloor + p_{m,i}^{(t)}$$
$$\bar{\rho}_{m,i}^{(t)} = \max\{0, \bar{q}_{m,i}^{(t)} - \bar{q}_{m,i}^{(t-1)}\}$$
13  **end**
14 **end**

**Algorithm 5:** The Complete Online Algorithm - *COA*

**Input:** $K, M, I, \beta, \delta, \mathbf{h}, \mathbf{b}, \mathbf{s}, \mathbf{z}, \mathbf{l}, \mathbf{d}^{in}, \mathbf{d}^{out}, \epsilon$
**Output:** $\bar{\mathbf{q}}, \bar{\rho}, \bar{\mathbf{x}}, \bar{\mathbf{y}}$

1 Initialization: $\mathbf{q} = \mathbf{0}, \mathbf{x} = \mathbf{0}, \mathbf{y} = \mathbf{0}, \bar{\mathbf{q}} = \mathbf{0}$;
2 call the local clustering algorithm in Alg. 2;
3 **for** *each time slot* $t \in [T]$ **do**
4  $(\mathbf{q}^{(t)}, \mathbf{x}^{(t)}, \mathbf{y}^{(t)}) = ORFA^{(t)}(K, M, I, \beta, \delta, \mathbf{h}, \mathbf{b}, \mathbf{s}, \mathbf{z}, \mathbf{l}, \mathbf{d}^{in}, \mathbf{d}^{out}, \epsilon)$;
5  $(\mathbf{w}^{(t)}, \mathbf{p}^{(t)}, A, B, \Upsilon, G(V_m^{(t)}, E_m^{(t)}), \forall m \in [M]) = INIT(K, M, I, \mathbf{q}^{(t)}, \mathbf{c}, \mathbf{b})$;
6  $(\bar{\mathbf{q}}^{(t)}, \bar{\rho}^{(t)}) = OWDR(\mathbf{q}^{(t)}, G(V_m^{(t)}, E_m^{(t)}), \forall m \in [M], \mathbf{p}^{(t)}, \mathbf{w}^{(t)}, \bar{\mathbf{q}}^{(t-1)})$;
7  Obtain $\bar{\mathbf{x}}^{(t)}$ and $\bar{\mathbf{y}}^{(t)}$ by solving LP (14) using interior point method;
8 **end**

*4) Flow Redirection:* $\bar{\mathbf{q}}^{(t)}$ and $\bar{\rho}^{(t)}$ produced by OWDR together with flow routing decisions $\mathbf{x}^{(t)}$ and $\mathbf{y}^{(t)}$ from Alg. 1 may well not be a feasible solution of (7). We next obtain $\bar{\mathbf{x}}^{(t)}$ and $\bar{\mathbf{y}}^{(t)}$ based on $(\bar{\mathbf{q}}^{(t)}, \bar{\rho}^{(t)})$ that leads to a complete feasible solution of (7), by solving a linear optimization problem as follows. The LP in (14) is derived from (7) by plugging in the values of $\bar{\mathbf{q}}^{(t)}$ and $\bar{\rho}^{(t)}$ obtained by OWDR and removing the resulting fixed *VNF running cost* and *VNF deployment cost* from the objective and the relevant constraints.

$$\text{minimize} \quad C_T^{(t)} + C_E^{(t)} \quad (14)$$

subject to: constraints (8a)(8c)-(8g)

The above LP can be exactly solved using the interior point method in polynomial time.

### B. The Complete Online Algorithm

We summarize our complete online algorithm in Alg. 5. Here $ORFA^{(t)}$ is referring to the steps executed in time slot $t$ of the online algorithm ORFA, *i.e.*, lines 3-6 in Alg. 1. $INIT$ is the routine in Alg. 3 and $OWDR$ is the routine in Alg. 4.

**Theorem 5.** *The complete online algorithm COA runs in polynomial time.*

**Theorem 6.** *The complete online algorithm COA in Alg. 5 computes a feasible solution $(\bar{\mathbf{q}}^{(t)}, \bar{\mathbf{x}}^{(t)}, \bar{\mathbf{y}}^{(t)}, \bar{\rho}^{(t)})$ of the original problem (7).*

*Proof.* Since $\bar{\mathbf{q}}^{(t)}$ satisfies the condition in Theorem 4, we can find a feasible solution of $\mathbf{y}^{(t)}$ which satisfies constraints (8a) and (8c), and then a feasible solution of $\mathbf{x}^{(t)}$ satisfying the flow conservation constraints (8d) and (8e). That is, we can find a feasible solution when solving LP (14), and $(\bar{\mathbf{q}}^{(t)}, \bar{\mathbf{x}}^{(t)}, \bar{\mathbf{y}}^{(t)}, \bar{\rho}^{(t)})$ is feasible for the original problem (7). □

solution $\bar{q}_{m,i}^{(t)}$ can be produced by the original integral part $\lfloor q_{m,i}^{(t)} \rfloor$ plus the rounded fractional part $p_{m,i}^{(t)}$. Furthermore, based on the integral solutions $\bar{q}_{m,i}^{(t)}$, we can then calculate $\bar{\rho}_{m,i}^{(t)}$ as $\max\{0, \bar{q}_{m,i}^{(t)} - \bar{q}_{m,i}^{(t-1)}\}$ (line 12).

The following theorem shows that the numbers of instances of each VNF $m$ to deploy in the datacenters, $\bar{q}_{m,i}^{(t)}$ produced by OWDR, are integers and are sufficient to serve all the incoming flows that need this VNF at $t$.

**Theorem 4.** *Elements of $\bar{\mathbf{q}}^{(t)}$ produced by OWDR are non-negative integers and satisfy $\sum_{i \in [I]} q_{m,i}^{(t)} b_{m,i} \geq \sum_{k \in [K]} \hat{F}_{k,m}^{(t)}, \forall m \in [M], t \in [T]$.*

The detailed proof is given in Appendix D.

**Competitive Analysis.** We now analyze the competitive ratio of COA. Let $P_I(ORFA)$ denote the objective value of problem (11) achieved by its best integer solution. Let $\bar{P}_I(COA)$ be the objective value of the original problem (7) achieved by the solution produced by COA, in which the VNF running cost, VNF deployment cost, flow transfer cost and delay cost are $\bar{P}_I(C_R), \bar{P}_I(C_D), \bar{P}_I(C_T)$ and $\bar{P}_I(C_E)$, respectively:

$$\bar{P}_I(COA) = \bar{P}_I(C_R) + \bar{P}_I(C_D) + \bar{P}_I(C_E) + \bar{P}_I(C_T)$$

**Lemma 1.** *$\bar{P}_I(C_R)$ is at most 2 times $P_I(ORFA)$.*

**Lemma 2.** *$\bar{P}_I(C_D)$ is no larger than $\phi_1$ times $P_I(ORFA)$, where $\phi_1 = \max_{t \in [T], m \in [M], i \in [I]} \frac{\delta_{m,i}}{c_{m,i}^{(t)}}$ is the maximal ratio of deployment cost to operational cost per VNF instance.*

**Lemma 3.** $\bar{P}_I(C_T)$ *is no larger than* $\phi_2$ *times* $P_I(ORFA)$, *where* $\phi_2 = \max_{t\in[T], m\in[M], i\in[I]} \frac{(d_i^{in}+d_i^{out})\cdot b_{m,i}}{c_{m,i}^{(t)}}$ *is the maximal ratio of flow transfer cost to operational cost per VNF instance.*

**Lemma 4.** $\bar{P}_I(C_E)$ *is no larger than* $\phi_3$ *times* $P_I(ORFA)$, *where* $\phi_3 = \max_{t\in[T], m\in[M], i\in[I], k\in[K], u,v\in S} \max\{\frac{\alpha l_{u,v} a_k^{(t)} b_{m,i}}{Rc_{m,i}^{(t)}},\}$, $S = \cup_{k\in[K]}\{s_k, z_k\} \cup [I]$, *and* $\alpha$ *is the coefficient for the relaxed triangle inequality defined in (5).*

The proofs of the lemmas are given in Appendices E, F, G and H.

**Theorem 7.** *(Final Competitive Ratio)* $\bar{P}_I(COA)$ *is no larger than* $[\log(1 + MI/\epsilon) + 2](2 + \phi_1 + \phi_2 + \phi_3)$ *times the offline minimum objective value* $P_I^*$ *of the original problem (7).*

*Proof.* According to Lemmas 1 – 4 and Theorem 3, we have

$$\bar{P}_I(COA) \leq (2 + \phi_1 + \phi_2 + \phi_3)P_I(OFRA)$$
$$\leq [\log(1 + MI/\epsilon) + 2](2 + \phi_1 + \phi_2 + \phi_3)P_I^*$$
□

## VI. PERFORMANCE EVALUATION

### A. Simulation Setup

We use the topology of the physical network of Cogent to create the datacenter network [33]. The default number of datacenters is 50. Delays between datacenters are set proportional to their geographic distances, perturbed by multiplying a random number in $[0.8, 1.2]$. The flows are generated according to real web traffic based on the Wikipedia trace [34], which contains 20.6 billion HTTP requests within a 10-month period. We further boost the daily peak in the WikiPedia traffic to flash crowds generated using the model in [35]: $R_{flash}$ = shock level $\times R_{normal}$, where the *shock level* is the order of magnitude increase in the average request rate. Fig. 3 shows the input traffic rate to service chains, produced at shock level of 5 (default). Since the IP addresses in the traces are hidden due to privacy issues, we use the demographic density data of the global Internet users [36] to model the geo-distribution of the resource/destination pairs. A location with high demographic density is selected with a high probability to be a flow source or destination.

We simulate the VNFs in the following table, with respective processing capacity, instance type (following those of Amazon EC2), and flow rate change ratios. The running cost, deployment cost and flow transfer cost for each VNF are set according to EC2's pricing of the respective instance type [25]. We set $a_k^{(t)}$'s for converting delay to cost to 1. We create 30 service chains, each containing 2~5 VNFs randomly chosen from the four. Each time slot in our experiments corresponds to one hour.

| VNF | CPU usage | Capacity | Instance Type | Change Rate $\beta$ |
|---|---|---|---|---|
| Firewall | 4 | 900Mbps | m4.xlarge | 0.8 − 1.0 |
| Proxy | 4 | 900Mbps | m4.xlarge | 1.0 |
| NAT | 2 | 900Mbps | m4.large | 1.0 |
| IDS | 8 | 600Mbps | m4.2xlarge | 0.8 − 1.0 |

### B. Performance of ORFA

We first examine the ratio achieved by ORFA in Alg. 1 without rounding, by dividing the overall cost achieved by the fractional solution of ORFA by the offline minimum overall cost. The offline minimal cost is obtained by solving (7) exactly using CVX and MOSEK Optimizer. Due to the complexity of solving the offline problem with a large number of variables, we set the default number of time slots to be $T = 200$.

Fig. 4 shows that the number of datacenters does not influence the result significantly. Under our setting, datacenters are residing in locations with high Internet user density with higher probability according to the reality. Fig. 5 further shows that the performance is quite stable with the increase of the total number of time slots that the system spans. The impact of $\epsilon$ is not obvious as shown in Fig. 4 and Fig. 5. The theoretical worst case ratio given in Theorem 2 is larger when the number of datacenters is larger and smaller when $\epsilon$ is larger; here we observe that in practice, their impact is not obvious. We set $\epsilon = 0.1$ in the following experiments.

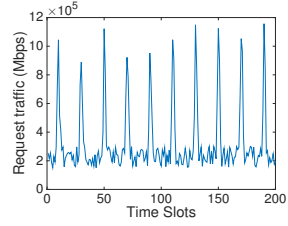

Fig. 3. Request traffic with flash crowds

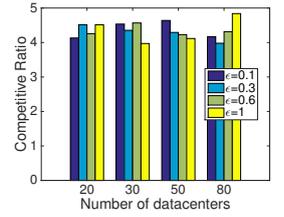

Fig. 4. Performance of ORFA with different # of datacenters

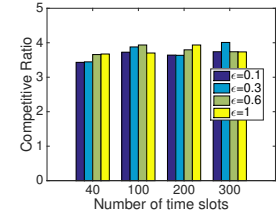

Fig. 5. Performance of ORFA with different # of time slots

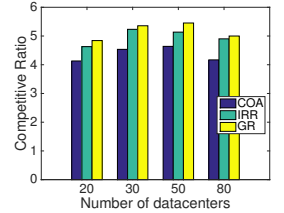

Fig. 6. Performance of COA with different # of datacenters

### C. Performance of COA

We next evaluate ratios achieved by our complete online algorithm, COA in Alg. 1, derived by dividing the overall cost achieved by the integer solution of COA by the offline minimum overall cost of (7). We compare the ratios achieved by our algorithm and those of two other algorithms: (i) *IRR*, our online algorithm ORFA in Alg. 1 together with a randomized *independent* rounding algorithm, which simply rounds the VNF quantities to nearest integers (*e.g.*, 1.4 to 1), and computes the ratio only if the result flow routing is feasible; and (ii) *GR*, ORFA with a greedy rounding algorithm which rounds up the fractional VNF quantities to guarantee feasibility. Fig. 6 and Fig. 7 show that our online algorithm with the dependent rounding approach consistently outperforms the other algorithms.

### D. Performance under Different Shock Levels

Fig 8 plots the ratios under different shock levels of the flash crowds. As expected, a higher shock level brings larger ratios. However, even when the shock level increases significantly from 1 to 100, our ratios only increase from about 1 to around 6, which shows the stability of the performance of our algorithms even in extreme scenarios.

## VII. CONCLUDING REMARKS

The fast adoption and development of Network Function Virtualization solutions introduce a series of challenges in modern datacenter management. This work aims to efficiently manage cloud resources for VNF deployment to realize the NFV goals of significant cost reduction and ease of management, and also guarantee short end-to-end delay experienced by the flows. We first leverage a regularization method from online learning to reshape the relaxation of the original offline optimization problem. Then the new problem can be decomposed into a set of sub-problems, each of which can be solved optimally in polynomial time. Furthermore, we design an online dependent rounding scheme to obtain the final randomized mixed integer solution. We show that the final solution to the original problem yields a competitive ratio which is independent of the number of flows or time horizon, based on the analysis via a primal-dual framework. The cost effectiveness is further validated by both theoretical proof and a series of trace-driven simulations.

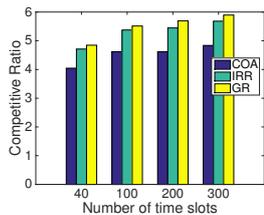

Fig. 7. Performance of COA with different # of time slots

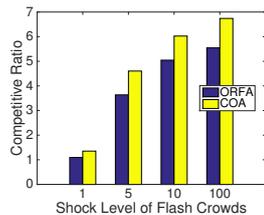

Fig. 8. Performance of our algorithms at various shock levels

## APPENDIX A
## PROOF OF THEOREM 2

*Proof. Basic Idea:* We derive the competitive ratio of *ORFA*, by comparing the overall cost it achieves (denoted by $P_f(OFRA)$) to the optimal overall cost of the relaxed offline problem $P_f$ (denoted by $P_f^*$) instead. Then because the optimal objective value of the relaxed problem $P_f$ is no larger than the optimum of the original offline problem in (7), we know the ratio we derived is also an upper bound of the ratio computed by comparing the overall cost *ORFA* achieves to the offline optimum of the original problem.

The competitive analysis is based on the primal-dual framework. We formulate the dual $D_f$ of $P_f$. Let $D_f^*$ denote the optimal (maximum) objective value of $D_f$. In order to obtain an upper bound of the ratio of $P_f(OFRA)$ to $P_f^*$, we need a lower bound of $P_f^*$. Since $P_f$ is an linear program for which strong duality holds [27], we have $P_f^* = D_f^*$, and seek to find a lower bound for $D_f^*$ instead. Note that the dual is a maximization problem that any feasible solution produces a lower bound of $D_f^*$, i.e., $D_f \leq D_f^* = P_f^*$ (here we abuse the notation $D_f$ to denote the objective value on a feasible solution of $D_f$). Therefore, the key is to identify a feasible solution of the dual.

We first construct the dual problem $D_f$ of $P_f$ as follows:

$$D_f : \text{maximize} \sum_{t \in [T]} \sum_{k \in [K]} \sum_{m \in [M]} \hat{F}_{k,m}^{(t)} \mu_{k,m}^{(t)} \quad (15)$$

subject to:

$$c_{m,i}^{(t)} - b_{m,i}\lambda_{m,i}^{(t)} + \nu_{m,i}^{(t)} - \nu_{m,i}^{(t+1)} \geq 0$$
$$\forall t \in [T], m \in [M], i \in [I] \tag{16a}$$

$$\delta_{m,i} - \nu_{m,i}^{(t)} \geq 0$$
$$\forall t \in [T], m \in [M], i \in [I] \tag{16b}$$

$$\lambda_{m,i}^{(t)} + (d_i^{in} + d_i^{out}\beta_{k,m}) + \xi_{k,m,i}^{(t)} - \mu_{k,m}^{(t)}$$
$$- \gamma_{k,m,i}^{(t)} + \beta_{k,m}\tau_{k,m,i}^{(t)} \geq 0$$
$$t \in [T], k \in [K], m \in [M]/\{m_{k,z}, m_{k,z}\}, i \in [I] \tag{16c}$$

$$\lambda_{m,i}^{(t)} + (d_i^{in} + d_i^{out}\beta_{k,m}) + \xi_{k,m,i}^{(t)} - \mu_{k,m}^{(t)} + \beta_{k,m}\tau_{k,m,i}^{(t)} \geq 0$$
$$t \in [T], k \in [K], m = m_{k,s}, i \in [I] \tag{16d}$$

$$\lambda_{m,i}^{(t)} + (d_i^{in} + d_i^{out}\beta_{k,m}) + \xi_{k,m,i}^{(t)} - \mu_{k,m}^{(t)} - \gamma_{k,m,i}^{(t)} \geq 0$$
$$t \in [T], k \in [K], m = m_{k,z}, i \in [I] \tag{16e}$$

$$\omega_{k,m,i,m',i'}^{(t)} + h_{k,m,m'}\gamma_{k,m',i'}^{(t)} - h_{k,m,m'}\tau_{k,m,i}^{(t)} \geq 0$$
$$\forall t \in [T], k \in [K], m \in [M], i \in [I], m' \in [M], i' \in [I]/\{i\} \tag{16f}$$

$$\omega_{k,m,i,m',i}^{(t)} + h_{k,m,m'}\gamma_{k,m',i}^{(t)} - h_{k,m,m'}\tau_{k,m,i}^{(t)}$$
$$- h_{k,m,m'}(d_i^{in} + d_i^{out}) \geq 0$$
$$\forall t \in [T], k \in [K], m \in [M], i \in [I], m' \in [M] \tag{16g}$$

By exploring the regularized problem $\tilde{P}_f$, we define $\tilde{D}_f$ as its dual problem. Let $\lambda_{m,i}^{(t)}, \mu_k^{(t)}, \gamma_{k,m,i}^{(t)}, \tau_{k,m,i}^{(t)}$ be dual variables of $\tilde{D}_f$ associated with constraints (8a),(8c),(8d) and (8e) respectively. The value of $q_f^{(t)*}$, $x_f^{(t)*}$ and $y_f^{(t)*}$ satisfy the K.K.T. Optimality Condition [27], which is sufficient and necessary for optimality of a convex optimization as follows:

$$q_{m,i}^{(t)} - \sum_{k\in[K]} y_{k,m,i}^{(t)} \geq 0$$
$$\lambda_{m,i}^{(t)}(q_{m,i}^{(t)} - \sum_{k\in[K]} y_{k,m,i}^{(t)}) = 0 \tag{17a}$$

$$c_{m,i}^{(t)} - b_{m,i}\lambda_{m,i}^{(t)} + \nu_{m,i}^{(t)} - \nu_{m,i}^{(t+1)} \geq 0$$
$$q_{m,i}^{(t)}(c_{m,i}^{(t)} - b_{m,i}\lambda_{m,i}^{(t)} + \nu_{m,i}^{(t)} - \nu_{m,i}^{(t+1)}) = 0 \tag{17b}$$

$$\delta_{m,i} - \nu_{m,i}^{(t)} \geq 0$$
$$\rho_{m,i}^{(t)}(\delta_{m,i} - \nu_{m,i}^{(t)}) = 0 \tag{17c}$$

$$\lambda_{m,i}^{(t)} + (d_i^{in} + d_i^{out}\beta_{k,m}) + \xi_{k,m,i}^{(t)} - \mu_{k,m}^{(t)}$$
$$- \gamma_{k,m,i}^{(t)} + \beta_{k,m}\tau_{k,m,i}^{(t)} \geq 0$$
$$y_{k,m,i}^{(t)}(\lambda_{m,i}^{(t)} + (d_i^{in} + d_i^{out}\beta_{k,m}) + \xi_{k,m,i}^{(t)} - \mu_{k,m}^{(t)}$$
$$- \gamma_{k,m,i}^{(t)} + \beta_{k,m}\tau_{k,m,i}^{(t)}) = 0 \tag{17d}$$

$$\lambda_{m,i}^{(t)} + (d_i^{in} + d_i^{out}\beta_{k,m}) + \xi_{k,m,i}^{(t)} - \mu_{k,m}^{(t)} + \beta_{k,m}\tau_{k,m,i}^{(t)} \geq 0$$
$$y_{k,m,i}^{(t)}(\lambda_{m,i}^{(t)} + (d_i^{in} + d_i^{out}\beta_{k,m}) + \xi_{k,m,i}^{(t)} - \mu_{k,m}^{(t)} + \beta_{k,m}\tau_{k,m,i}^{(t)})$$
$$= 0 \tag{17e}$$

$$\lambda_{m,i}^{(t)} + (d_i^{in} + d_i^{out}\beta_{k,m}) + \xi_{k,m,i}^{(t)} - \mu_{k,m}^{(t)} - \gamma_{k,m,i}^{(t)} \geq 0$$
$$y_{k,m,i}^{(t)}(\lambda_{m,i}^{(t)} + (d_i^{in} + d_i^{out}\beta_{k,m}) + \xi_{k,m,i}^{(t)} - \mu_{k,m}^{(t)} - \gamma_{k,m,i}^{(t)}) = 0 \tag{17f}$$

$$\omega_{k,m,i,m',i'}^{(t)} + h_{k,m,m'}\gamma_{k,m',i'}^{(t)} - h_{k,m,m'}\tau_{k,m,i}^{(t)} \geq 0$$
$$x_{k,m,i,m',i'}^{(t)}(\omega_{k,m,i,m',i'}^{(t)} + h_{k,m,m'}\gamma_{k,m',i'}^{(t)} - h_{k,m,m'}\tau_{k,m,i}^{(t)}) = 0 \tag{17g}$$

$$\omega_{k,m,i,m',i}^{(t)} + h_{k,m,m'}\gamma_{k,m',i}^{(t)} - h_{k,m,m'}\tau_{k,m,i}^{(t)}$$
$$- h_{k,m,m'}(d_i^{in} + d_i^{out}) \geq 0$$
$$x_{k,m,i,m',i}^{(t)}(\omega_{k,m,i,m',i}^{(t)} + h_{k,m,m'}\gamma_{k,m',i}^{(t)} - h_{k,m,m'}\tau_{k,m,i}^{(t)}$$
$$- h_{k,m,m'}(d_i^{in} + d_i^{out})) = 0 \tag{17h}$$

Define $n = MI$ as a global notation. Directed by the K.K.T Optimality Condition (17a)-(17h), a feasible solution of the regularized dual problem can be derived as follows:

$$\dot{\lambda}_{m,i}^{(t)} = \tilde{\lambda}_{m,i}^{(t)*}, \dot{\mu}_k^{(t)} = \mu_k^{(t)*}, \dot{\gamma}_{k,m,i}^{(t)} = \gamma_{k,m,i}^{(t)*}, \dot{\tau}_{k,m,i}^{(t)} = \tau_{k,m,i}^{(t)*} \tag{18}$$

$$\dot{\nu}_{m,i}^{(t)} = \frac{\delta_{m,i}}{\eta} \ln\left(\frac{1 + \frac{\epsilon}{n}}{\tilde{q}_{m,i}^{(t)*} + \frac{\epsilon}{n}}\right) \tag{19}$$

In the following, to be clear, we use $\dot{D}_f$ to denote the dual problem constructed in (18) and (19).

**Lemma 5.** $\sum_{t\in[T]} \sum_{k\in[K]} \sum_{m\in[M]} \hat{F}_{k,m}^{(t)} \tilde{\mu}_{k,m}^{(t)}$ *is a lower bound of* $\tilde{P}_f^*$.

*Proof.* According to (18) and (19), $\dot{D}_f$ is obtained by plugging each $\dot{\mu}_{k,m}^{(t)}$ into the objective function of (15). We will further show that (18) and (19) are feasible to (15). Given $\tilde{q}_{m,i}^{(t)*} \geq 0$: Choose $\eta = \ln(1 + \frac{n}{\epsilon})$ and plug it into (19), we have:

$$\nu_{m,i}^{(t)} = \frac{\delta_{m,i}}{\ln(1 + \frac{n}{\epsilon})} \ln\left(\frac{1 + \frac{\epsilon}{n}}{q_{m,i}^{(t-1)} + \frac{\epsilon}{n}}\right) \tag{20}$$

$$0 \leq \nu_{m,i}^{(t)} \leq \delta_{m,i} \tag{21}$$

By (19) and (17b), we also have:

$$c_{m,i}^{(t)} - b_{m,i}\lambda_{m,i}^{(t)} - \frac{\delta_{m,i}}{\eta}\ln(\frac{q_{m,i}^{(t)} + \frac{\epsilon}{n}}{q_{m,i}^{(t-1)} + \frac{\epsilon}{n}}) \geq 0$$
$$q_{m,i}^{(t)}\left(c_{m,i}^{(t)} - b_{m,i}\lambda_{m,i}^{(t)} - \frac{\delta_{m,i}}{\eta}\ln(\frac{q_{m,i}^{(t)} + \frac{\epsilon}{n}}{q_{m,i}^{(t-1)} + \frac{\epsilon}{n}})\right) = 0 \tag{22}$$

Which shows (16b) is feasible, and also:

$$b_{m,i}\lambda_{m,i}^{(t)} + \nu_{m,i}^{(t)} - \nu_{m,i}^{(t-1)}$$
$$= b_{m,i}\lambda_{m,i}^{(t)} - \frac{\delta_{m,i}}{\eta} \ln\left(\frac{q_{m,i}^{(t)} + \frac{\epsilon}{n}}{q_{m,i}^{(t-1)} + \frac{\epsilon}{n}}\right) \tag{23}$$
$$\leq c_{m,i}^{(t)}$$

which indicates (16a) is feasible, thus $\dot{D}_f \leq D_f^* \leq D_f^*$ according to weak duality [27]. Since $\sum_{t\in[T]} \sum_{k\in[K]} \sum_{m\in[M]} \hat{F}_{k,m}^{(t)} \tilde{\mu}_{k,m}^{(t)}$ is the objective value of $\dot{D}_f$, shows the feasibility of ORFA.

Compare the total cost of fractional VNF deployment problem using online algorithm ORFA with the offline optimal algorithm. The total cost in (7) can be divided into total VNF running cost, total VNF deployment cost, total flow transfer cost and total end-to-end cost. We compute the ratio of each of them to $D_f$, the competitive ratio can be bounded by summing up the four sub-ratios. In the following, we simplify $\tilde{\psi}^*$ to $\psi$ where $\psi$ generally denotes all the variables. □

First give the bound on the VNF deployment cost $P_f(C_D)$ and the sum of the VNF Running Cost $P_f(C_R)$, the Flow Transfer Cost $P_f(C_T)$ and the End-to-End Cost $P_f(C_E)$ in the following two lemmas, respectively.

**Lemma 6.** *The VNF Deployment Cost of $P_f(C_D)$ is no larger than $\dot{D}_f(\ln(1 + n/\epsilon) + \frac{1}{\phi})$, where $\phi$ is the minimum positive* *fractional deployment for every VNF in each datacenter such that* $\phi = \min_{t\in[T], i\in[I], m\in[M]} q_{m,i}^{(t)} \quad \forall q_{m,i}^{(t)} > 0$.

**Lemma 7.** *(Bounding the sum of other cost)* The sum of VNF Running Cost, Flow Transfer Cost and End-to-End Cost $P_f(C_R) + P_f(C_T) + P_f(C_E)$ is no larger than $\dot{D}_f$.

Based on the bound for the VNF Deployment Cost $P_f(C_D)$ in Lemma 6 and the sum of the VNF Running Cost $P_f(C_R)$, the Flow Transfer Cost $P_f(C_T)$ and the End-to-End Cost $P_f(C_E)$ in Lemma 7, we have:

$$\frac{P_f(ORFA)}{P_f^*} \le \ln(1+\frac{n}{\epsilon}) + 1 + \frac{1}{\phi} \quad (24)$$

Since $P_f^*$ is the optimum of the relaxation of $P^*$, hence:

$$P_f(ORFA) \le \left(\ln(1+\frac{n}{\epsilon})+1+\frac{1}{\phi}\right)P_f^* \le \left(\ln(1+\frac{n}{\epsilon})+1+\frac{1}{\phi}\right)P^* \quad (25)$$
$\square$

## APPENDIX B
## PROOF OF LEMMA 6

*Proof.* We define $\dot{D}_f^{(t)}$ as the objective value of total cost during time slot $t \in [T]$, and also $P_f^{(t)}(C_\Psi)$ as correspondent cost of $P_f(C_\Psi)$ during time slot $t \in [T]$, e.g. $P_f^{(t)}(C_D)$ is the VNF Deployment Cost during time slot $t \in [T]$, such that:

$$\dot{D}_f^{(t)} = P_f^{(t)}(C_D) + P_f^{(t)}(C_R) + P_f^{(t)}(C_T) + P_f^{(t)}(C_E) \quad (26)$$

The value of $\dot{D}_f^{(t)}$ can be calculated by the objective function of the dual problem given from (15) such that:

$$\dot{D}_f^{(t)} = \sum_{k\in[K]} \sum_{m\in[M]} \hat{F}_{k,m}^{(t)} \mu_{k,m}^{(t)} \quad (27)$$

Derived from (4), calculate the value of $P_f^{(t)}(C_D)$:

$$P_f^{(t)}(C_D) = \sum_{m\in[M]} \sum_{i\in[I]} \rho_{m,i}^{(t)} \delta_{m,i} \quad (28a)$$

$$= \eta \sum_{q_{m,i}^{(t)} > q_{m,i}^{(t-1)}} \frac{\delta_{m,i}}{\eta}(q_{m,i}^{(t)} - q_{m,i}^{(t-1)})$$

$$\le \eta \sum_{q_{m,i}^{(t)} > q_{m,i}^{(t-1)}} (q_{m,i}^{(t)} + \frac{\epsilon}{n})(\frac{\delta_{m,i}}{\eta} \ln(\frac{q_{m,i}^{(t)} + \frac{\epsilon}{n}}{q_{m,i}^{(t-1)} + \frac{\epsilon}{n}})) \quad (28b)$$

$$= \eta \sum_{q_{m,i}^{(t)} > q_{m,i}^{(t-1)}} q_{m,i}^{(t)}(\frac{\delta_{m,i}}{\eta} \ln(\frac{q_{m,i}^{(t)} + \frac{\epsilon}{n}}{q_{m,i}^{(t-1)} + \frac{\epsilon}{n}}))$$

$$+ \eta \frac{\epsilon}{n} \sum_{q_{m,i}^{(t)} > q_{m,i}^{(t-1)}} (\delta_{m,i} \ln(\frac{q_{m,i}^{(t)} + \frac{\epsilon}{n}}{q_{m,i}^{(t-1)} + \frac{\epsilon}{n}})) \quad (28c)$$

$$= P_f^{(t)}(C_D') + P_f^{(t)}(C_D'') \quad (28d)$$

(28b) is derived from the inequality as follows:
$$a - b \le a\ln(a/b) \quad \forall a,b \in \mathbb{R}^+$$

and (28d) is derived from the definitions as follows:

$$P_f^{(t)}(C_D') \triangleq \eta \sum_{q_{m,i}^{(t)} > q_{m,i}^{(t-1)}} q_{m,i}^{(t)} \left(\frac{\delta_{m,i}}{\eta} \ln(\frac{q_{m,i}^{(t)} + \frac{\epsilon}{n}}{q_{m,i}^{(t-1)} + \frac{\epsilon}{n}})\right)$$

$$P_f^{(t)}(C_D'') \triangleq \eta \frac{\epsilon}{n} \sum_{q_{m,i}^{(t)} > q_{m,i}^{(t-1)}} q_{m,i}^{(t)} \left(\frac{\delta_{m,i}}{\eta} \ln(\frac{q_{m,i}^{(t)} + \frac{\epsilon}{n}}{q_{m,i}^{(t-1)} + \frac{\epsilon}{n}})\right)$$

We next prove:
$$P_f^{(t)}(C_D') \le \dot{D}_f^{(t)} \ln(1 + n/\epsilon) \quad (29)$$

as follows:

$$P_f^{(t)}(C_D') = \eta \sum_{q_{m,i}^{(t)} > q_{m,i}^{(t-1)}} q_{m,i}^{(t)}(b_{m,i}\lambda_{m,i}^{(t)} - c_{m,i}^{(t)}) \quad (30a)$$

$$\le \eta \sum_{m\in[M]} \sum_{i\in[I]} q_{m,i}^{(t)} b_{m,i}\lambda_{m,i}^{(t)} \quad (30b)$$

$$= \eta \sum_{m\in[M]} \sum_{i\in[I]} \lambda_{m,i}^{(t)} \sum_{k\in[K]} y_{k,m,i}^{(t)} \quad (30c)$$

$$= \eta \sum_{k\in[K]} \sum_{m\in[M]} \mu_{k,m}^{(t)} \hat{F}_{k,m}^{(t)} - \eta \sum_{k\in[K]} \sum_{m\in[M]} \sum_{i\in[I]} \xi_{k,m,i}^{(t)} y_{k,m,i}^{(t)}$$

$$- \eta \sum_{k\in[K]} \sum_{m\in[M]} \sum_{i\in[I]} (d_i^{in} + d_i^{out}\beta_{k,m}) y_{k,m,i}^{(t)}$$

$$- \eta(-\sum_{k\in[K]} \sum_{m\in[M]/\{m_{k,s}\}} \sum_{i\in[I]} \gamma_{k,m,i}^{(t)} y_{k,m,i}^{(t)})$$

$$- \eta(\sum_{k\in[K]} \sum_{m\in[M]/\{m_{k,z}\}} \sum_{i\in[I]} \beta_{k,m}\tau_{k,m,i}^{(t)} y_{k,m,i}^{(t)}) \quad (30d)$$

(30a) is derived from (22), (30d) is derived from (17d), (17e), (17f), (8c) and KKT conditions (22) (17a). The following proofs will also be useful in the proofs of Lemma 7 that:

$$P_f^{(t)}(C_T) + P_f^{(t)}(C_E) =$$
$$\sum_{k\in[K]} \sum_{m\in[M]} \sum_{i\in[I]} (d_i^{in} + d_i^{out}\beta_{k,m}) y_{k,m,i}^{(t)}$$
$$+ \sum_{k\in[K]} \sum_{m\in[M]} \sum_{i\in[I]} \xi_{k,m,i}^{(t)} y_{k,m,i}^{(t)}$$
$$- \sum_{k\in[K]} \sum_{m\in[M]/\{m_{k,s}\}} \sum_{i\in[I]} \gamma_{k,m,i}^{(t)} y_{k,m,i}^{(t)}$$
$$+ \sum_{k\in[K]} \sum_{m\in[M]/\{m_{k,z}\}} \sum_{i\in[I]} \beta_{k,m}\tau_{k,m,i}^{(t)} y_{k,m,i}^{(t)} \quad (31)$$

We first compute:

$$\sum_{k\in[K]} \sum_{m\in[M]/\{m_{k,z}\}} \sum_{i\in[I]} \beta_{k,m}\tau_{k,m,i}^{(t)} y_{k,m,i}^{(t)}$$
$$- \sum_{k\in[K]} \sum_{m\in[M]/\{m_{k,s}\}} \sum_{i\in[I]} \gamma_{k,m,i}^{(t)} y_{k,m,i}^{(t)} \quad (32a)$$

$$= \sum_{k\in[K]} \sum_{m\in[M]/\{m_{k,z}\}} \sum_{i\in[I]} \sum_{m'\in[M]} \sum_{i'\in[I]} \tau_{k,m,i}^{(t)} h_{k,m,m'} x_{k,m,i,m',i'}^{(t)}$$
$$- \sum_{k\in[K]} \sum_{m\in[M]/\{m_{k,s}\}} \sum_{i\in[I]} \sum_{m'\in[M]} \sum_{i'\in[I]} \gamma_{k,m,i}^{(t)} h_{k,m',m} x_{k,m',i',m,i}^{(t)} \quad (32b)$$

$$= \sum_{k\in[K]} \sum_{m\in[M]} \sum_{i\in[I]} \sum_{m'\in[M]} \sum_{i'\in[I]} \tau_{k,m,i}^{(t)} h_{k,m,m'} x_{k,m,i,m',i'}^{(t)}$$
$$- \sum_{k\in[K]} \sum_{m\in[M]} \sum_{i\in[I]} \sum_{m'\in[M]} \sum_{i'\in[I]} \gamma_{k,m,i}^{(t)} h_{k,m',m} x_{k,m',i',m,i}^{(t)} \quad (32c)$$

$$= \sum_{k\in[K]} \sum_{m\in[M]} \sum_{i\in[I]} \sum_{m'\in[M]} \sum_{i'\in[I]} \omega_{k,m,i,m',i'}^{(t)} x_{k,m,i,m',i'}^{(t)}$$
$$- \sum_{k\in[K]} \sum_{m\in[M]} \sum_{i\in[I]} \sum_{m'\in[M]} h_{k,m,m'}(d_i^{in} + d_i^{out}) x_{k,m,i,m',i'}^{(t)} \quad (32d)$$

(32a) is derived from the flow conservation constraints (8d) and (8e), (32c) holds since $h_{k,m,m'}$ is always 0 if $m$ is the last VNF in flow $k$'s service chain that $m = m_{k,z}$, whereas $h_{k,m',m}$ is always 0 if $m$ is the first VNF in flow $k$'s service chain that $m = m_{k,s}$. (32d) is derived from (17g) and (17h).

By the proofs from (32a)-(32d), equation (31) holds. Then by (30a)-(30d),(27) and (31), we get:

$$P_f^{(t)}(C_D') \le \eta \dot{D}_f^{(t)} - \eta\left(P_f^{(t)}(C_T) + P_f^{(t)}(C_E)\right) \le \eta \dot{D}_f^{(t)}$$

Since $\eta = \ln(1+n/\epsilon)$, (29) holds.

Then we bound the value of $P_f^{(t)}(C_D'')$, by the definition $\phi = \min_{t\in[T],i\in[I],m\in[M]} q_{m,i}^{(t)} \quad \forall q_{m,i}^{(t)} > 0$:

$$P_f^{(t)}(C_D') \ge \phi \sum_{q_{m,i}^{(t)} > q_{m,i}^{(t-1)}} (\delta_{m,i} \ln(\frac{q_{m,i}^{(t)} + \frac{\epsilon}{n}}{q_{m,i}^{(t-1)} + \frac{\epsilon}{n}})) = \frac{n\phi}{\epsilon} P_f^{(t)}(C_D'')$$

Then by (29):

$$P_f^{(t)}(C_D'') \le \frac{\epsilon}{n\phi} P_f^{(t)}(C_D') \le \frac{1}{\phi}\frac{\eta\epsilon}{n}\dot{D}_f \le \frac{1}{\phi}\dot{D}_f^{(t)}$$

Note that $\frac{\eta\epsilon}{n} \le 1$ holds since $\ln(1+x) > x, x \in \mathbb{R}^+$ and $\eta = \ln(1+n/\epsilon)$

Finally, we get the bound for the *VNF Deployment Cost* that:

$$P_f^{(t)}(C_D) = P_f^{(t)}(C_D') + P_f^{(t)}(C_D'') \le (\ln(1+n/\epsilon) + \frac{1}{\phi})\dot{D}_f^{(t)} \quad (33)$$

Then take the summation of value $P_f^{(t)}(C_D)$ and $\dot{D}_f^{(t)}$ over time slot $t \in [T]$ from (33), we get:

$$P_f(C_D) \le (\ln(1+n/\epsilon) + \frac{1}{\phi})\dot{D}_f$$

□

## APPENDIX C
## PROOF OF THE LEMMA 7

*Proof.* The VNF Running Cost by definition (2) is:

$$P_f^{(t)}(C_R) = \sum_{m\in[M]} \sum_{i\in[I]} c_{m,i}^{(t)} q_{m,i}^{(t)} \quad (34a)$$

$$= \sum_{m\in[M]} \sum_{i\in[I]} q_{m,i}^{(t)} \left(b_{m,i}\lambda_{m,i}^{(t)} + \frac{\delta_{m,i}}{\eta}\ln(\frac{q_{m,i}^{(t)} + \frac{\epsilon}{n}}{q_{m,i}^{(t-1)} + \frac{\epsilon}{n}})\right) \quad (34b)$$

$$\le \sum_{m\in[M]} \sum_{i\in[I]} q_{m,i}^{(t)} b_{m,i} \lambda_{m,i}^{(t)} \quad (34c)$$

$$= \sum_{m\in[M]} \sum_{i\in[I]} \lambda_{m,i}^{(t)} \sum_{k\in[K]} y_{k,m,i}^{(t)} \quad (34d)$$

$$= \sum_{k\in[K]} \sum_{m\in[M]} \mu_{k,m}^{(t)} \hat{F}_{k,m}^{(t)}$$
$$- \sum_{k\in[K]} \sum_{m\in[M]} \sum_{i\in[I]} (d_i^{in} + d_i^{out}\beta_{k,m}) y_{k,m,i}^{(t)}$$
$$- \sum_{k\in[K]} \sum_{m\in[M]} \sum_{i\in[I]} \xi_{k,m,i}^{(t)} y_{k,m,i}^{(t)} \quad (34e)$$
$$+ \sum_{k\in[K]} \sum_{m\in[M]/\{m_{k,s}\}} \sum_{i\in[I]} \gamma_{k,m,i}^{(t)} y_{k,m,i}^{(t)})$$
$$- \eta(\sum_{k\in[K]} \sum_{m\in[M]/\{m_{k,z}\}} \sum_{i\in[I]} \beta_{k,m}\tau_{k,m,i}^{(t)} y_{k,m,i}^{(t)})$$

(34b) is derived from (22), (34d) is derived from (17a), (34e) is derived from (17d), (17e),(17f), (8c) and (22).

Together with the results from (34e),(27) and (31) we have:

$$P_f^{(t)}(C_R) + P_f^{(t)}(C_T) + P_f^{(t)}(C_E) \le \dot{D}_f^{(t)} \quad (35)$$

Then take the summation of value $P_f^{(t)}(C_R)$, $P_f^{(t)}(C_T)$, $P_f^{(t)}(C_E)$ and $\dot{D}_f^{(t)}$ over time slot $t \in [T]$ from (36), we have:

$$P_f(C_R) + P_f(C_T) + P_f(C_E) \le \dot{D}_f \quad (36)$$

□

## APPENDIX D
## PROOF OF THEOREM 4

We first prove the following three lemmas. Where Lemma 8 is a necessary condition for feasible solutions of (7),9 and Lemma 10 are two important properties for the rounded solution by COA [7].

**Lemma 8.** $\mathbf{q}^{(t)}$ *is a feasible solution of VNF deployment for the original problem (7) if the following inequality (37) holds:*

$$\forall t \in [T], m \in [M]$$
$$\sum_{i\in[I]} q_{m,i}^{(t)} b_{m,i} \ge \sum_{k\in[K]} \hat{F}_{k,m}^{(t)} \quad (37)$$

*Proof.* If (37) holds, then we can find a feasible solution of $\mathbf{y}^{(t)}$ satisfies (8a) and (8c). Then by (8d) and (8e) we can also get feasible solution $\mathbf{x}^{(t)}$ of the original problem (7), thus $\mathbf{q}^{(t)}$ is a feasible solution of VNF deployment for (7). □

**Lemma 9.** *The rounded solution of $\bar{\mathbf{q}}^{(t)}$ satisfies the* Marginal Distribution *property (38) defined as follows:*

$$\forall t \in [T], m \in [M], i \in [I]$$
$$\Pr(\bar{q}_{m,i}^{(t)} = \lfloor q_{m,i}^{(t)} \rfloor + 1) = q_{m,i}^{(t)} - \lfloor q_{m,i}^{(t)} \rfloor \quad (38)$$
$$\Pr(\bar{q}_{m,i}^{(t)} = \lfloor q_{m,i}^{(t)} \rfloor) = \lfloor q_{m,i}^{(t)} \rfloor + 1 - q_{m,i}^{(t)}$$

The dependent rounding algorithm follows the *Marginal Distribution* property, which is a basic property of the traditional rounding algorithms and it also holds for the dependent rounding algorithm, the original detailed proof is long and can be found in [7]. The proof shows that during each iteration of the dependent rounding algorithm, the expectation of value $q_{m,i}^{(t)}$ remains the same and always follows the *Marginal Distribution* property.

**Lemma 10.** *The solution of COA satisfies the* Weighed Degree-preservation *Property: denote the fractional weighed degree of the buffer data center as $d_{m,j_m^{(t)*}}^{(t)} = \sum_{i\in[I]/\{p_{m,j_m^{(t)*}}^{(t)}\}} w_{m,i}^{(t)} p_{m,i}^{(t)}$. We have:*

$$\bar{q}_{m,j_m^{(t)*}}^{(t)} = \lceil d_{m,j_m^{(t)*}}^{(t)} + q_{m,j_m^{(t)*}}^{(t)} \rceil \quad (39)$$

*with probability exactly 1 in any iteration of Alg. 4.*

*Proof.* To show the proof of Lemma 10, first we fix any buffer datacenter $j_m^{(t)*}$ with fractional weighted degree:

$$d_{m,j_m^{(t)*}}^{(t)} = \sum_{i\in[I]/\{j_m^{(t)*}\}} w_{m,i}^{(t)} p_{m,i}^{(t)}$$

If $j_m^{(t)*}$ has at most one floating edge incident on it at the beginning of OWDR, it is easy to verify that (39) holds; so suppose $j_m^{(t)*}$ initially had at least two floating edges incident on it. We claim that as long as i has at least two floating edges incident on it, and we start with iterations that has path with length 2. Note that during each iteration the value of $d_{m,j_m^{(t)*}}^{(t)}$ remain the same since one edge is rounded down that decrease the value of $d_{m,j_m^{(t)*}}^{(t)}$ while the other edges rounded up which will increase the same value of $d_{m,j_m^{(t)*}}^{(t)}$ by the definition of $w_{m,i}^{(t)}$. Revisit OWDR, each iteration eliminate at least one floating from each cluster, such we will finally get the scenario that at most one floating edge remained and during all the iterations the equality (39) always hold, therefore the *Weighed Degree-preservation* Property holds for OWDR. □

Back to Theorem 4, first note that (37) holds before OWDR since the fractional solution from regularization guarantee the feasibility of (7) by Theorem (1). During each iteration of OWDR, the value of of $\sum_{i\in[I]} q_{m,i}^{(t)} b_{m,i}$ remains the same the by the property of Lemma 10. Therefore the the condition (37) always holds, so our rounded solution $\bar{\mathbf{q}}^{(t)}$ by OWDR is feasible for the original problem (7).

# APPENDIX E
## PROOF OF LEMMA 1

Define $(\mathbf{q}^{(t)*}, \mathbf{x}^{(t)*}, \mathbf{y}^{(t)*}, \rho^{(t)*})$ as the best solution for (7). $(\mathbf{q}^{(t)}, \mathbf{x}^{(t)}, \mathbf{y}^{(t)}, \rho^{(t)})$ is the best fractional solution for problem (11).

*Proof.* For the proof of Lemma 1, we first prove $\forall t \in [T]$:

$$\sum_{m \in [M]} \sum_{i \in [I]} c_{m,i}^{(t)} \bar{q}_{m,i}^{(t)} \leq 2 \sum_{m \in [M]} \sum_{i \in [I]} c_{m,i}^{(t)} \tilde{q}_{m,i}^{(t)} \quad (40)$$

By the *marginal distribution* property of the OWDR algorithm from Lemma 9, we have

$$\sum_{m \in [M]} \sum_{i \in [I]/\{j_m^{(t)*}\}} c_{m,i}^{(t)} \bar{q}_{m,i}^{(t)} = \sum_{m \in [M]} \sum_{i \in [I]/\{j_m^{(t)*}\}} c_{m,i}^{(t)} q_{m,i}^{(t)} \quad (41)$$

for all the non-buffer datacenters. For the buffer datacenter:

$$\begin{aligned}
\sum_{m \in [M]} c_{m,j_m^{(t)*}}^{(t)} \bar{q}_{m,j_m^{(t)*}}^{(t)} &\leq \sum_{m \in [M]} c_{m,j_m^{(t)*}}^{(t)} \Big( q_{m,j^{(t)*}}^{(t)} \\
&+ \sum_{i \in A_{m,j^{(t)*}}} \{q_{m,i}^{(t)}\}(1-\{q_{m,i}^{(t)}\}) \frac{b_{m,i}}{b_{m,j^*}} \Big) \\
&\leq \sum_{m \in [M]} c_{m,j_m^{(t)*}}^{(t)} q_{m,j_m^{(t)*}}^{(t)} + \sum_{m \in [M]} \sum_{i \in [I]/\{j_m^{(t)*}\}} c_{m,j_m^{(t)*}}^{(t)} \\
& q_{m,i}^{(t)} \frac{b_{m,i}}{b_{m,j^*}} \\
&\leq \sum_{m \in [M]} c_{m,j_m^{(t)*}}^{(t)} q_{m,j_m^{(t)*}}^{(t)} + \sum_{m \in [M]} \sum_{i \in [I]/\{j_m^{(t)*}\}} c_{m,i}^{(t)} q_{m,i}^{(t)} \\
&= \sum_{m \in [M]} \sum_{i \in [I]} c_{m,i}^{(t)} q_{m,i}^{(t)}
\end{aligned} \quad (42)$$

By (41) and (42), (40) holds. And since:

$$\sum_{m \in [M]} \sum_{i \in [I]/\{j_m^{(t)*}\}} c_{m,i}^{(t)} q_{m,i}^{(t)} \leq P_I(ORFA)$$

Therefore Lemma 1 holds. □

# APPENDIX F
## PROOF OF LEMMA 2

*Proof.* By the definition $\phi_1 = \max_{t \in [T], m \in [M], i \in [I]} \frac{\delta_{m,i}}{c_{m,i}^{(t)}}$

$$\begin{aligned}
\bar{P}_I(C_D) &= \sum_{t \in [T]} \sum_{m \in [M]} \sum_{i \in [I]} \delta_{m,i} \bar{\rho}_{m,i}^{(t)} \\
&= \sum_{t \in [T]} \sum_{m \in [M]} \sum_{i \in [I]} \frac{\delta_{m,i}}{c_{m,i}^{(t)}} c_{m,i}^{(t)} \bar{\rho}_{m,i}^{(t)} \\
&\leq \sum_{t \in [T]} \sum_{m \in [M]} \sum_{i \in [I]} \phi_1 c_{m,i}^{(t)} \bar{q}_{m,i}^{(t)} = \phi_1 \bar{P}_I(C_R)
\end{aligned}$$

By Lemma 1 and the definition of $\bar{P}_I(C_D)$, Lemma 2 holds. □

# APPENDIX G
## PROOF OF LEMMA 3

*Proof.* By the definition
$\phi_2 = \max_{t \in [T], m \in [M], i \in [I]} \frac{(d_i^{in}+d_i^{out})b_{m,i}}{c_{m,i}^{(t)}}$:

$$\begin{aligned}
\bar{P}_I(C_T) &= \sum_{t \in [T]} \sum_{k \in [K]} \sum_{m \in [M]} \sum_{i \in [I]} (d_i^{in}+d_i^{out}\beta_{k,m}) y_{k,m,i}^{(t)} - \\
&\sum_{t \in [T]} \sum_{k \in [K]} \sum_{m \in [M]} \sum_{m' \in [M]} \sum_{i \in [I]} h_{k,m,m'}(d_i^{in}+d_i^{out}) x_{k,m,i,m',i}^{(t)} \\
&\leq \sum_{t \in [T]} \sum_{k \in [K]} \sum_{m \in [M]} \sum_{i \in [I]} (d_i^{in}+d_i^{out}\beta_{k,m}) y_{k,m,i}^{(t)} \\
&\leq \phi_2 \sum_{t \in [T]} \sum_{m \in [M]} \sum_{i \in [I]} c_{m,i}^{(t)} \sum_{k \in [K]} \frac{y_{k,m,i}^{(t)}}{b_{m,i}} \\
&\leq \phi_2 \sum_{t \in [T]} \sum_{m \in [M]} \sum_{i \in [I]} c_{m,i}^{(t)} q_{m,i}^{(t)} = \phi_2 \bar{P}_I(C_R)
\end{aligned}$$

By Lemma 1 and the definition of $\bar{P}_I(C_T)$, Lemma 3 holds. □

# APPENDIX H
## PROOF OF LEMMA 4

*Proof.* By the definition
$\phi_3 = \max_{t \in [T], m \in [M], i \in [I], k \in [K], u,v \in S} \max\{\frac{\alpha l_{u,v} a_k^{(t)} b_{m,i}}{R c_{m,i}^{(t)}},\}$, $S = \cup_{k \in [K]} \{s_k, z_k\} \cup [I]$:

$$\begin{aligned}
\bar{P}_I(C_E) &= \sum_{t \in [T]} \sum_{k \in [K]} \sum_{m \in [M]} \sum_{i \in [I]} \xi_{k,m,i}^{(t)} y_{k,m,i}^{(t)} \\
&+ \sum_{k \in [K]} \sum_{m \in [M]} \sum_{i \in [I]} \sum_{m' \in [M]} \sum_{i' \in [I]} \omega_{k,m,i,m',i'}^{(t)} x_{k,m,i,m',i'}^{(t)} \\
&\leq \phi_3 \sum_{t \in [T]} \sum_{k \in [K]} \sum_{m \in [M]/\{m_{k,s}, m_{k,z}\}} \sum_{i \in [I]} c_{m,i}^{(t)} \frac{y_{k,m,i}^{(t)}}{b_{m,i}} + \\
&\phi_3 \sum_{t \in [T]} \sum_{k \in [K]} \sum_{m \in \{m_{k,s}, m_{k,z}\}} \sum_{i \in [I]} c_{m,i}^{(t)} \frac{y_{k,m,i}^{(t)}}{b_{m,i}} \\
&= \phi_3 \sum_{t \in [T]} \sum_{m \in [M]} \sum_{i \in [I]} c_{m,i}^{(t)} \sum_{k \in [K]} \frac{y_{k,m,i}^{(t)}}{b_{m,i}} \\
&\leq \phi_3 \sum_{t \in [T]} \sum_{m \in [M]} \sum_{i \in [I]} c_{m,i}^{(t)} q_{m,i}^{(t)} = \phi_3 \bar{P}_I(C_R)
\end{aligned}$$

By Lemma 1 and the definition of $\bar{P}_I(C_E)$, Lemma 4 holds. □